\documentclass[aps,footinbib, notitlepage,superscriptaddress, longbibliography,eqsecnum]{revtex4-1}
\usepackage{graphicx}
 \usepackage{amssymb}
 \usepackage{amsmath}
 \usepackage{amsfonts}
\usepackage{overpic}
\usepackage{accents}
\usepackage{enumerate}
\usepackage{color}
\usepackage[dvipsnames]{xcolor}
\usepackage{xspace}
\usepackage[normalsize]{subfigure}
\usepackage{hyperref}
\usepackage{bm}
\usepackage{empheq}
\usepackage[capitalize]{cleveref}
\crefname{section}{Sec.}{Secs.}

\DeclareFontFamily{OT1}{pzc}{}
\DeclareFontShape{OT1}{pzc}{m}{it}{<-> s * [1.10] pzcmi7t}{}
\DeclareMathAlphabet{\mathpzc}{OT1}{pzc}{m}{it}

\providecommand{\st}[1]{_{\text{#1}}}
\providecommand{\sfrac}[2]{#1/#2}

\providecommand{\ut}[1]{^{\text{#1}}}

\def\onehalf{\frac{1}{2}}

\def\bra{\ensuremath{\langle}}
\def\ket{\ensuremath{\rangle}}
\def\eq{\st{st}}
\def\eqBlk{\st{b,st}}

\def\pd{\partial}

\def\im{\mathrm{i}}

\def\qv{\bv{q}}

\def\Jv{\bv{J}}

\def\nv{\bv{n}}

\def\bvnp{\bv{0}_\parallel}

\def\pv{\bv{p}}

\def\rv{\bv{r}}
\def\rvp{\bv{r}_\parallel}

\def\Rv{\bv{R}}

\def\b0{\bv{0}}

\def\eff{\st{eff}}

\def\Fcal{\mathcal{F}}
\def\Gcal{\mathcal{G}}
\def\Hcal{\mathcal{H}}
\def\Hc2{\Hcal^{(2)}}
\def\Ical{\mathcal{I}}
\def\Ccal{\mathcal{C}}
\def\Dcal{\mathcal{D}}
\def\Ecal{\mathcal{E}}
\def\Kcal{\mathcal{K}}
\def\Lcal{\mathcal{L}}
\def\Mcal{\mathcal{M}}
\def\Ncal{\mathcal{N}}
\def\mcal{\mathpzc{m}}
\def\Ocal{\mathcal{O}}
\def\Pcal{\mathcal{P}}
\def\Qcal{\mathcal{Q}}

\def\Tcal{\mathcal{T}}
\def\Ucal{\mathcal{U}}

\def\Wcal{\mathcal{W}}
\def\Zcal{\mathcal{Z}}

\def\wcal{\mathpzc{w}}

\def\hyp13{{_1 F_3}}

\def\reg{\text{reg}}

\def\bcs{BCs\xspace}
\def\pbc{\ut{(p)}}

\def\Dbc{\ut{(D)}}
\def\Nbc{\ut{(N)}}

\def\izero{^{(0)}}

\def\tred{\ring{\tau}}
\def\tauLG{\tau}

\def\amplXip{\xi_+\izero}

\def\d{\mathrm{d}}

\def\cutoff{\varepsilon}
\def\tphys{{\tilde t}}
\def\GAMMA{\Gamma}

\newcommand{\bitem}{\begin{itemize}}
\newcommand{\eitem}{\end{itemize}}
\newcommand{\benum}{\begin{enumerate}}
\newcommand{\eenum}{\end{enumerate}}
\newcommand{\btab}[1]{\begin{tabular}{#1}}
\newcommand{\etab}{\end{tabular}}
\newcommand{\beq}{\begin{equation}}
\newcommand{\eeq}{\end{equation}}
\newcommand{\beqn}{\begin{equation*}}
\newcommand{\eeqn}{\end{equation*}}

\newcommand{\bv}[1]{\mathbf{#1}}

\graphicspath{{Figs/}{figs/}{}}

\begin{document}
\title{Fluctuations of the critical Casimir force}
\author{Markus Gross}
\email{gross@is.mpg.de}
\affiliation{Max-Planck-Institut f\"{u}r Intelligente Systeme, Heisenbergstra{\ss}e 3, 70569 Stuttgart, Germany}
\affiliation{IV.\ Institut f\"{u}r Theoretische Physik, Universit\"{a}t Stuttgart, Pfaffenwaldring 57, 70569 Stuttgart, Germany}
\author{Andrea Gambassi}
\affiliation{SISSA -- International School for Advanced Studies and INFN, via Bonomea 265, 34136 Trieste, Italy}
\author{S. Dietrich}
\affiliation{Max-Planck-Institut f\"{u}r Intelligente Systeme, Heisenbergstra{\ss}e 3, 70569 Stuttgart, Germany}
\affiliation{IV.\ Institut f\"{u}r Theoretische Physik, Universit\"{a}t Stuttgart, Pfaffenwaldring 57, 70569 Stuttgart, Germany}
\date{\today}

\begin{abstract}
The critical Casimir force (CCF) arises from confining fluctuations in a critical fluid and thus it is a fluctuating quantity itself. While the mean CCF is universal, its (static) variance has previously been found to depend on the microscopic details of the system which effectively set a large-momentum cutoff in the underlying field theory, rendering it potentially large. 
This raises the question how the properties of the force variance are reflected in experimentally observable quantities, such as the thickness of a wetting film or the position of a suspended colloidal particle.
Here, based on a rigorous definition of the instantaneous force, we analyze static and dynamic correlations of the CCF for a conserved fluid in film geometry for various boundary conditions within the Gaussian approximation.  
We find that the dynamic correlation function of the CCF is independent of the momentum cutoff and decays algebraically in time. Within the Gaussian approximation, the associated exponent depends only on the dynamic universality class but not on the boundary conditions.
We furthermore consider a fluid film, the thickness of which can fluctuate under the influence of the time-dependent CCF. 
The latter gives rise to an effective non-Markovian noise in the equation of motion of the film boundary and induces a distinct contribution to the position variance. 
Within the approximations used here, at short times, this contribution grows algebraically in time whereas, at long times, it saturates and contributes to the steady-state variance of the film thickness.
\end{abstract}

%\pacs{68.03.Kn, 05.40.-a, 47.11.-j, 47.35.Pq, 83.80.Fg}

\maketitle

\section{Introduction}

A fluid close to a continuous phase transition exhibits remarkable universal properties, such as long-ranged fluctuations and scale invariance \cite{le_bellac_quantum_1991}.
A core element of criticality is the notion of an order parameter (OP) $\phi$, which, e.g., in the case of a binary liquid mixture is proportional to the deviation of the concentration $c_A$ of species A from its critical value $c_{A,c}$, i.e., $\phi\propto c_A-c_{A,c}$.
Confining a critical fluid leads to a critical Casimir force (CCF) acting on the confining boundaries \cite{fisher_wall_1978,krech_casimir_1994,brankov_theory_2000}.
Such a situation arises, e.g., when two colloidal particles immersed in a critical solvent come close to another or if one colloidal particle approaches a container wall of the sample.
CCFs thus provide means to control the aggregation behavior of colloidal suspensions \cite{hertlein_direct_2008} and are, accordingly, not only of theoretical but also of highly practical interest.
Consequently, CCFs and, more generally, Casimir-like fluctuation-induced forces have been extensively studied in as well as out of equilibrium (see, e.g., Refs. \cite{gambassi_casimir_2009,furukawa_nonequilibrium_2013,maciolek_collective_2018,gross_dynamics_2019,rohwer_correlations_2019,callegari_optical_2021} and references therein).

Previous studies mainly focused on the mean value of the thermally averaged CCF, which is finite and universal, i.e., it does not depend on the specific material considered, but only on the bulk and surface universality class \footnotetext[100]{This notion of universality applies to a fluid formed by particles with short-ranged interactions. In the presence of van der Waals forces, in particular, the CCF can exhibit a non-universal behavior \cite{dantchev_universality_2003,dantchev_interplay_2007,valchev_sign_2017}} \cite{diehl_field-theoretical_1986,Note100}.
However, since the CCF arises from fluctuations of the OP, it is itself a fluctuating quantity.
In Ref.\ \cite{bartolo_fluctuations_2002} the equilibrium fluctuations of the CCF in a generic Gaussian medium subject to Dirichlet \bcs have been investigated theoretically, and it has been shown that the static equilibrium variance depends in a non-universal way on the microscopic details of the model (specifically, on the microscopic length scale below which the continuum description breaks down). 
These findings were confirmed analytically and via Monte-Carlo simulations of lattice models for the case of periodic \bcs \cite{dantchev_critical_2004}. 
The variance of the CCF acting on membrane inclusions has been found to be strongly cutoff dependent as well \cite{bitbol_fluctuations_2010}.
In order to understand these results, we note that the (instantaneous) CCF $\Kcal$ (per area) is given by the difference between the pressures acting on the two sides of a boundary, $\Kcal=\Pcal_f - \Pcal_b$, where the subscripts refer to \emph{f}ilm and \emph{b}ulk (a notion to be specified below; see also \cref{fig_sketch}). 
Concerning the mean value $\bra\Kcal\ket$, all cutoff-dependent quantities present in $\Pcal_{f,b}$ cancel in the difference, giving rise to a finite and in fact universal quantity. 
In contrast, the variance $\bra (\Delta\Kcal)^2\ket = \bra\Delta\Pcal_f^2\ket + \bra\Delta\Pcal_b^2\ket$ (where $\Delta \Kcal \equiv \Kcal-\bra \Kcal\ket$) is the sum of the individual variances of $\Pcal_f$ and $\Pcal_b$ (assuming statistical independence of film and bulk) and thus no cancellation of divergences can occur \cite{bartolo_fluctuations_2002}.

Notably, the detailed form of the cutoff dependence of the variance implies a divergence in the strict continuum limit.
The corresponding huge variances predicted in this way for the CCF seemingly stand in contrast to the fact that experiments on CCFs in wetting films \cite{garcia_critical_1999, ganshin_critical_2006, fukuto_critical_2005, rafai_repulsive_2007} or in colloidal systems \cite{hertlein_direct_2008, gambassi_critical_2009} did not observe such giant fluctuations.
Colloidal particles in critical bulk fluids rather show essentially standard Brownian diffusion with an effective diffusivity modified due to the fluctuations of the field \cite{demery_perturbative_2011, dean_diffusion_2011, demery_diffusion_2013}. 
Moreover, near a wall, colloids diffusively explore the corresponding effective potential induced by the mean CCF \cite{paladugu_nonadditivity_2016, magazzu_controlling_2019}.

Divergent equilibrium variances of the force have also been obtained for the quantum-electro\-dynamical Casimir effect \cite{barton_fluctuations_1991}.
However, it has been shown that the divergence problem is less severe or even absent when considering temporally averaged quantities \cite{barton_fluctuations_1991,bartolo_fluctuations_2002}. In fact, such a temporal average is indispensable due to the necessarily finite temporal resolution of a measurement apparatus.
This issue can also be appreciated in view of a Langevin description of a Brownian particle \cite{gardiner_stochastic_2009}: while the random force has correlations proportional to a Dirac $\delta$ function and thus is strongly divergent, the resulting observable quantities, such as the position distribution of the particle, are finite \cite{demery_perturbative_2011, dean_diffusion_2011, demery_diffusion_2013}.
We finally remark that the static variance of thermal van der Waals forces between dielectric slabs has been found to be finite and independent of a microscopic cutoff, without the necessity for temporal averaging \cite{dean_fluctuation_2013}.

In the present study, based on the non-equilibrium stress tensor formalism \cite{dean_out--equilibrium_2010,kruger_stresses_2018,gross_surface-induced_2018,gross_dynamics_2019}, we investigate static and dynamic equilibrium fluctuations of the CCF in film geometry for various non-symmetry breaking \bcs. 
We focus on the Gaussian approximation. Simulations of lattice models \cite{dantchev_critical_2004} have shown that it constitutes an accurate description of the (static) stress probability distribution.
In accordance with Refs.\ \cite{bartolo_fluctuations_2002,dantchev_critical_2004} we confirm the generic non-universal and divergent character of the \emph{static} variance of the CCF acting on a fixed boundary.
Going beyond previous studies, we show that the \emph{dynamic} correlations of the CCF are finite and decay algebraically in time with exponents depending only on the spatial dimensionality and the dynamic universality class.
We furthermore consider a movable film boundary subject to the fluctuations of the CCF. 
We find that the mean-squared displacement of the boundary is, at short times, characterized by an algebraic growth (with possible logarithmic corrections), while, at long times, it saturates at a finite value due to the confining potential stemming from the CCF.
Accordingly, the strong divergences of the static variance of the CCF as function of the cutoff do not show up in its dynamic correlations or in experimentally observable quantities such as the position of the film boundary.

We present our study as follows: In \cref{sec_prelim}, we introduce the model and briefly review the formalism required to determine time-dependent CCFs. 
In \cref{sec_fluct_CCF}, we analyze static and dynamic fluctuations of the CCF in the case of a film with fixed boundaries.
This constraint is released in \cref{sec_fluct_boundary}, where we consider the thermal fluctuations of the film thickness.
A summary of the findings of the present study is provided in \cref{sec_summ}.

\section{Preliminaries}
\label{sec_prelim}

\subsection{Order-parameter dynamics}
\label{sec_OP_dyn}

In thermal equilibrium at temperature $T$ [defined here in units of $k_B$, i.e., $T=T\st{phys}/(1/k_B)$] and in $d$ spatial dimensions, the OP $\phi$ follows the probability distribution 
\beq P\st{eq}[\phi] \sim e^{-\sfrac{\Fcal[\phi]}{T}},
\label{eq_eq_dist}\eeq 
with the Hamiltonian  ($\rv = (r_1,\ldots, r_d)$)
\begin{multline}  \Fcal[\phi] \equiv  \int_V \d^d r\, \Hcal(\rv,L, \phi(\rv), \nabla\phi(\rv), \tauLG) ,\quad  \Hcal(\rv,L, \phi(\rv), \nabla\phi(\rv), \tauLG) = \Hcal_b(\phi(\rv), \nabla\phi(\rv), \tauLG) + \Hcal_s(\rv,L, \phi(\rv), \nabla\phi(\rv));
\label{eq_freeEn}\end{multline} 
accordingly, $\Fcal$ and $T$ have the dimension energy and $\Fcal/T$ is dimensionless.
For the bulk Hamiltonian density $\Hcal_b$ we consider a Gaussian Landau-Ginzburg form,
\beq \Hcal_b(\phi,\nabla\phi, \tauLG) = \onehalf (\nabla \phi)^2 + \frac{\tauLG}{2} \phi^2, %+ \frac{g}{4!} \phi^4
\label{eq_Hamiltonian}\eeq 
with $(\nabla\phi)^2 = \sum_{\alpha=1}^d (\pd_\alpha\phi)^2$, $\pd_\alpha = \frac{\pd}{\pd r_\alpha}\equiv \nabla_\alpha$, and the temperature parameter 
\beq \tauLG =(\amplXip)^{-2} \tred,
\label{eq_tauLGDef}
\eeq 
which is directly proportional to the reduced temperature
\beq \tred \equiv \frac{T-T_{c}}{T_{c}} > 0,
\label{eq_T_red}\eeq 
where $T_c$ denotes the (bulk) critical temperature. 
Furthermore, $\amplXip$ denotes a nonuniversal critical amplitude defined via the correlation length $\xi$ of the fluctuations as 
\beq \xi(\tred\to 0^+) = \amplXip \tred^{-\nu}= (\amplXip)^{1-2\nu} \tauLG^{-\nu},
\label{eq_correl_len}\eeq 
with a bulk critical exponent $\nu$.
Within the Gaussian approximation, one has $\nu=1/2$, such that \cref{eq_correl_len} reduces to $\xi = 1/\sqrt{\tauLG}$. For typical simple fluids one finds $\amplXip \approx\, 1\,\mathrm{nm}$ \cite{fukuto_critical_2005}.
For the surface Hamiltonian density $\Hcal_s$ we do not assume a specific form, except that it is strongly localized at a boundary; e.g., for a planar surface at $z=L$, it takes the form 
\beq \Hcal_s(\rv,L,\phi,\nabla\phi) = \delta(z-L)U_s(\phi,\nabla\phi)
\label{eq_Hden_s}\eeq 
with a potential $U_s$. 
We consider the thin film geometry with macroscopically large lateral directions and, accordingly, we decompose the $d$-dimensional vector $\rv$ as $\rv=\{\rvp=(r_1,\ldots, r_{d-1}),z\}$.
The following \bcs of the OP are considered in the transverse direction $z$:
\begin{subequations}
\begin{align}
    \text{periodic}: \qquad & \phi(\rvp,z) = \phi(\rvp,z+L), \\
    \text{Dirichlet}: \qquad & \phi(\rvp,0) = 0 = \phi(\rvp,L), \intertext{and}
    \text{Neumann}: \qquad & \pd_z\phi(\rvp,z)\big|_{z=0} = 0 = \pd_z\phi(\rvp,z)\big|_{z=L}.
\end{align}\label{eq_BCs}
\end{subequations}

Denoting the physical time by $\tphys$ [to be distinguished from the rescaled one, see \cref{eq_time_resc} below], the OP dynamics is taken to be described by model B \cite{hohenberg_theory_1977}:
\beq \pd_\tphys\phi = \gamma \nabla^2 \mu(\phi) + \bar\zeta,
\label{eq_modelB_bare}\eeq 
where $\gamma$ is a mobility coefficient and 
\beq \mu(\phi) \equiv \frac{\delta \Fcal(\tauLG; [\phi])}{\delta \phi} = \frac{\pd\Hcal}{\pd\phi} - \sum_{j=1}^d \nabla_j\left(\frac{\pd\Hcal}{\pd\nabla_j\phi}\right) \label{eq_chempot}\eeq 
is the chemical potential. 
Far away from a boundary, one has $\mu = -\nabla^2 \phi + \tauLG\phi $.
Furthermore, $\bar\zeta$ is a Gaussian white noise with zero mean and correlations
$\bra \bar\zeta(\rv, \tphys) \bar\zeta(\rv', \tphys\,')\ket = -2 T \gamma\, \nabla^2 \delta(\rv-\rv')\delta(\tphys-\tphys\,')
$.
The form of \cref{eq_modelB_bare} and of the noise ensure local conservation of the OP in an unconfined system. 
In order to guarantee local and global OP conservation in the presence of boundaries, the flux $\bm{J}=-\gamma\nabla\mu + \bm{\Ncal}$ of the OP (with noise $\bm{\Ncal}$ such that $\nabla\cdot\bm{\Ncal}=\bar\zeta$) must vanish at the confining surfaces of the system, i.e., $J_z=0$ for $z=0,L$ \footnote{In fact, we require the deterministic and stochastic fluxes, $\Jv_d=-\gamma\nabla\mu$ and $\Jv_s=\bm{\Ncal}$, to both vanish independently at the boundaries.}. 
This is ensured by periodic or Neumann \bcs, which we shall henceforth use in the dynamic model \footnote{Standard Dirichlet \bcs do not constrain the derivatives of $\phi$ at the boundary and therefore entail in general a non-vanishing flux $\Jv$ across a boundary. Thus they are not considered here; see Refs.\ \cite{diehl_boundary_1992,gross_dynamics_2019} for further discussion.}.
In order to simplify the notation, we follow Ref.\ \cite{gross_dynamics_2019} and remove the temperature $T$ from the description by introducing a rescaled OP field $\phi/T^{1/2}$ and, additionally, introduce a rescaled time
\beq t = \gamma \tphys,
\label{eq_time_resc}\eeq
having the dimension $[t] \sim [L]^4$.
Accordingly, \cref{eq_modelB_bare} turns into
\beq \pd_t \phi = -\nabla^4 \phi + \tauLG \nabla^2 \phi + \zeta,
\label{eq_modelB_resc}\eeq 
with a Gaussian noise $\zeta \equiv \bar\zeta/T^{1/2}\gamma$ correlated as
\beq \bra \zeta(\rv, t) \zeta(\rv', t')\ket = -2 \, \nabla^2 \delta(\rv-\rv')\delta(t-t').
\label{eq_noise_correl}\eeq

\begin{figure}[t]\centering
    \includegraphics[width=0.22\linewidth]{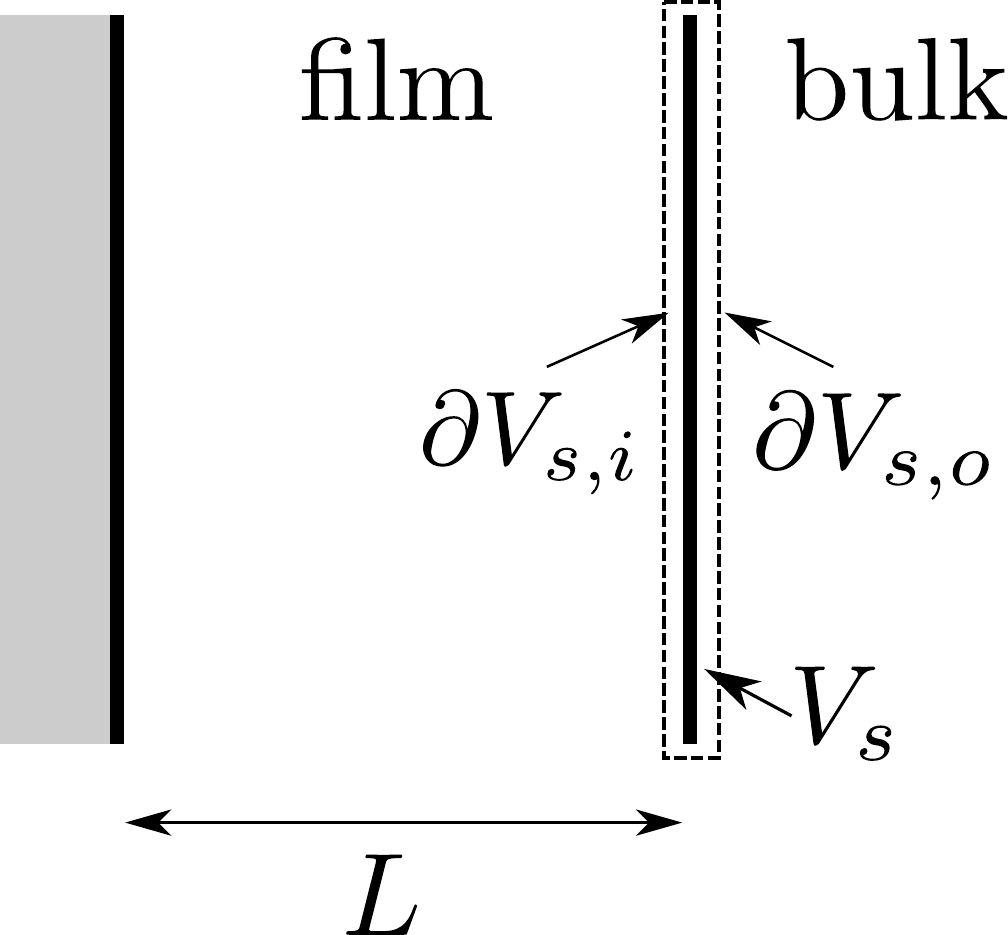}
    \caption{Sketch of the situation considered in the present study. A film of thickness $L$ is enclosed by two boundaries (thick black lines) each of surface area $A$ (spanning the plane perpendicular to the figure). The right boundary is enclosed by a thin imaginary volume $V_s$. The CCF acting on the right boundary is obtained from an integration of the stress tensor over the surface of $V_s$, which consists of an inner ($\pd V_{s,i}$) and an outer part ($\pd V_{s,o}$) [see \cref{eq_CCF_Pfilm}]. The film and the bulk consist of the same liquid whereas the gray area corresponds to a solid.}
    \label{fig_sketch}
\end{figure}

\subsection{Forces}
\label{sec_forces}

The instantaneous CCF alluded to in the introduction is defined in terms of a dynamic stress tensor which can be derived for generic Landau-Ginzburg-type of models  \cite{dean_out--equilibrium_2010}. The associated formalism has been developed further in Ref.\ \cite{kruger_stresses_2018} and has been applied in Refs.\ \cite{gross_surface-induced_2018,gross_dynamics_2019} for the case of a quench of a conserved fluid at criticality. 
Notably, the expressions for the stress tensor in and out of equilibrium are, in general, different \footnote{An exception is the case of Dirichlet \bcs, for which the time-dependent CCF in film geometry has been studied in Refs.\ \cite{gambassi_critical_2006,gambassi_relaxation_2008}}.
The concept of a dynamic stress tensor allows one to rigorously define an CCF at any instant during the time evolution of the system. 

The determination of time-dependent forces on boundaries within model B is detailed in Refs.\ \cite{gross_surface-induced_2018,gross_dynamics_2019}; in the following we provide only the essential equations. 
The dynamical stress tensor is given by \cite{gross_surface-induced_2018} 
\beq \bar \Tcal_{\alpha\beta} \equiv \Tcal_{\alpha\beta} + \mu\phi \delta_{\alpha\beta},
\label{eq_stressten_dyn}\eeq 
which is expressed here in terms of the standard (grand canonical) stress tensor
\beq \Tcal_{\alpha\beta} = \frac{\pd \Hcal}{\pd\nabla_\alpha\phi} \nabla_\beta \phi - \delta_{\alpha\beta} \Hcal.
\label{eq_stressten_gc}\eeq 
The last term in \cref{eq_stressten_dyn} involving the chemical potential $\mu$ [\cref{eq_chempot}] generically arises in a non-equilibrium situation for a system with no-flux \bcs.
Considering the right boundary of the film at position $z=L$ as shown in \cref{fig_sketch}, the instantaneous generalized force $\Kcal_\alpha$ (per area $A$) in direction $\alpha$ of the surface is defined as
\beq \Kcal_\alpha \equiv -\frac{1}{A}\frac{\pd \Fcal}{\pd L},
\label{eq_genforce}\eeq 
which can be transformed into \cite{gross_surface-induced_2018}
\beq \Kcal_\alpha  A = -\int_{\pd V_s} \d^{d-1} s_\beta\, \bar \Tcal_{\alpha\beta} + \int_{V_s} \d^d r\, (\nabla_\alpha \mu)\phi ,
\label{eq_CCF_stress}\eeq
where $V_s$ is an infinitesimally thin volume enclosing the surface. 
If $\phi$ fulfills no-flux \bcs or Dirichlet \bcs, the last term in \cref{eq_CCF_stress} vanishes.
We note that, due to the localized nature of $\Hcal_s$ [\cref{eq_Hden_s}], effectively only the bulk Hamiltonian density $\Hcal_b$ enters into $\bar\Tcal_{\alpha\beta}$ in \cref{eq_CCF_stress}.
In the case of periodic \bcs, \cref{eq_CCF_stress} has to be suitably symmetrized (see Ref.\ \cite{gross_dynamics_2019}). Since the resulting expression for the CCF is identical to omitting the last term in \cref{eq_CCF_stress}, we do not separately deal with this aspect.
In general, the instantaneous CCF (per area) $\Kcal\equiv \Kcal_z$ acting on a boundary at position $z$ of the film is obtained as
\beq  \Kcal(z,t) = \Pcal_f(z,t) - \Pcal_b(z,t) ,
\label{eq_CCF_Pfilm}\eeq 
where 
\beq \Pcal_{f,b}(z,t)  =  \frac{1}{A} \int_{\pd V_{s,i/o}} \d s_z(\rvp,z)\,  \bar\Tcal_{zz}(\rvp, z, t) 
\label{eq_Pfilm}\eeq 
represents the film and the bulk pressure; $\pd V_{s,i/o}$, correspondingly, denotes the bounding surface of the inner/outer fluid and $\rvp$ is the location of the $(d-1)$-dimensional surface element $\d s_z$. We emphasize that the quantity $\Kcal$ is actually a pressure. However, to stay in line with the convention employed in most of the literature, we use the term Casimir \emph{force} throughout this study.
In equilibrium, we assume the boundaries of the film to be fixed at $z\in \{0,L\}$. 
However, in order to facilitate later extensions of the model, in our formalism we shall keep $z$ general.
For the Landau-Ginzburg Hamiltonian density of \cref{eq_Hamiltonian} and upon using \cref{eq_chempot}, the dynamical stress tensor takes the form
\beq  \bar \Tcal_{zz} = \onehalf (\pd_z \phi)^2 - \onehalf \sum_{\alpha=1}^{d-1} (\pd_\alpha \phi)^2  - \phi \nabla^2\phi + \onehalf\tauLG\phi^2 ,
\label{eq_Tzz_dyn}\eeq 
which, as noted after \cref{eq_CCF_stress}, depends only on bulk terms.
The mean CCF (per area) acting on a surface at location $z$ follows from \cref{eq_CCF_Pfilm} as
\begin{subequations}
\begin{align}
\bra \Kcal(z,t)\ket &= \bra \Pcal_f(z) - \Pcal_b(z) \ket \equiv \frac{1}{A} \int_{\pd V_{s,i}} \d s_z(\rvp,z)\, \bra \bar T_{zz}(\rvp,z,t)\ket  - \text{\{bulk\}} \\
&= \Bigg[\onehalf \bra (\pd_z \phi(\rvp, z, t) )^2\ket - \onehalf \sum_{\alpha=1}^{d-1} \bra (\pd_\alpha \phi(\rvp, z, t))^2\ket  - \Big\bra \phi(\rvp, z, t) \sum_{\alpha=1}^{d} \pd_\alpha^2 \phi(\rvp, z, t) \Big\ket \Bigg]_{{\rvp=0}\atop{z=0,L}}   - \text{\{bulk\}}, \label{eq_CCF_correlb}
\end{align}\label{eq_CCF_Tzz}
\end{subequations}
\hspace{-0.17cm}where ``\{bulk\}'' denotes the corresponding expression evaluated in the bulk, which is obtained by integrating over $\pd V_{s,o}$ and calculating statistical averages with the corresponding bulk distribution of $\phi$. 

\subsection{Order-parameter correlation functions}
\label{sec_OPcorrel}

Here, we summarize some useful expressions for and properties of the equilibrium two-point two-time OP correlation function $\Ccal(\rv,\rv',t)\equiv \bra \phi(\rv,t) \phi(\rv',0)\ket = \Ccal(\rvp-\rvp',z,z',t)$, where the latter equality follows from translational invariance in the lateral directions.
The time-independent, static limit of $\Ccal$ is denoted by $\Ccal\eq$ and follows as $\Ccal\eq(\rvp,z,z') = \Ccal(\rvp,z,z,t\to 0)$.

\subsubsection{Bulk correlation function}

The equilibrium bulk correlation function for model B is given by (see, e.g., Ref.\ \cite{gross_dynamics_2019})
\beq 
\Ccal_b(r, t) \equiv \Ccal_b(\rv, t) = \int \frac{\d^{d} q}{(2\pi)^{d}}\, e^{\im \qv \cdot \rv}  \frac{e^{-\qv^2(\qv^2 + \tauLG) t}}{\qv^2 + \tauLG},
\label{eq_Cdyn_blk}
\eeq
which, as indicated by the notation, is spatially isotropic \footnote{Note that, due to the time-reversal symmetry of the model, all correlation functions in fact depend on $|t|$ instead of $t$.}.
In the static equilibrium case ($t=0$) for $d>2$ and at bulk criticality ($\tauLG=0$), one has 
\beq \Ccal\eqBlk(r) = \frac{\Gamma(d/2-1)}{4\pi^{d/2}\, r^{d-2}} ,\qquad (d>2).
\label{eq_Ceq_blk}\eeq 
In $d=2$ dimensions, the static bulk correlation function diverges logarithmically for large $r$ \cite{gross_dynamics_2019}.

We now analyze $\Ccal_b$ [\cref{eq_Cdyn_blk}] for $d>2$ in certain relevant limits.
Due to the exponential in the expression for $\Ccal_b$, the integrand in \cref{eq_Cdyn_blk} gives significant contributions only if $\qv^4 t + \qv^2\tauLG t\lesssim 1$. There are two characteristic asymptotic regimes depending on $\qv^2 \ll \tauLG$ or $\qv^2\gg \tauLG$, as can be seen by splitting the integral in \cref{eq_Cdyn_blk} accordingly:
\beq \begin{split} \Ccal_b(r, t)  &\simeq \frac{1}{\tauLG} \int_{|\qv|\ll \sqrt{\tauLG}} \frac{\d^{d} q}{(2\pi)^{d}}\, e^{\im \qv \cdot \rv - \qv^2 \tauLG t}  + \int_{|\qv|\gg \sqrt{\tauLG}} \frac{\d^{d} q}{(2\pi)^{d}}\, \qv^{-2} e^{\im \qv \cdot \rv - \qv^4 t} \\
&= t^{-d/2} \tauLG^{-d/2-1} \int_{|\hat\qv|\ll \tauLG t^{1/2}} \frac{\d^d\hat q}{(2\pi)^d}\, e^{\im \hat \qv\cdot\rv / \sqrt{\tauLG t} - \hat\qv^2} + t^{1/2-d/4} \int_{|\hat\qv|\gg \sqrt{\tauLG t^{1/2}}} \frac{\d^d\hat q}{(2\pi)^d} \hat\qv^{-2} e^{\im\hat\qv\cdot\rv/t^{1/4} - \hat\qv^4 }. 
\end{split}\label{eq_Cdyn_blk_asympt_split}\eeq 
If $\tauLG t^{1/2}\gg 1$, i.e., for long times $t\gg \tauLG^{-2}$, the second term on the r.h.s.\ of \cref{eq_Cdyn_blk_asympt_split} is exponentially suppressed, and we can obtain an asymptotic estimate of $\Ccal_b$ by extending the integration range in the first term to all $\qv$:
\beq\begin{split} \Ccal_b(r,t)\big|_{t\gg \tauLG^{-2}} &\simeq t^{-d/2} \tauLG^{-d/2-1}  \int\frac{\d^d \hat q}{(2\pi)^d} e^{\im \qv\cdot\rv/(\tauLG t)^{1/2}  -\hat\qv^2} \\
&= (4\pi)^{-d/2} t^{-d/2} \tauLG^{-d/2-1} \exp\left(-\frac{r^2}{4\tauLG t}\right).
\end{split}\label{eq_Cdyn_larget}\eeq 
On the other hand, for $\tauLG t^{1/2}\ll 1$, i.e., at short times $t\ll \tauLG^{-2}$, the second term in \cref{eq_Cdyn_blk_asympt_split} dominates, which can be shown as follows: for $\tauLG t^{1/2}\ll 1$, the first term in \cref{eq_Cdyn_blk_asympt_split} can be approximated as $t^{-d/2} \tauLG^{-d/2-1} \int \frac{\d^d \hat q}{(2\pi)^d} \exp(\im\hat\qv\cdot\rv/\sqrt{\tauLG t}) \simeq t^{-d/2} \tauLG^{-d/2-1} \sqrt{\tauLG t}\, \tilde\delta^{(d)}(\rv) = \tauLG^{-1}$, where $\tilde\delta^{(d)}$ stands for an approximation of the Dirac $\delta$ function. The ratio of the first and the second term follows as $(\tauLG^2 t)^{d/4-1/2}/\tauLG^{d/2}$. Accordingly, for $d>2$, the first term is indeed negligible at short times. The bulk correlation function can be asymptotically estimated by extending the integration in the second term in \cref{eq_Cdyn_blk_asympt_split} over the whole range of $\hat\qv$, resulting in \cite{gross_dynamics_2019}
\beq \begin{split} \Ccal_b(r,t)\big|_{\tauLG=0} &= t^{1/2-d/4} \int \frac{\d^d \hat q}{(2\pi)^d} \hat \qv^{-2} e^{\im \hat \qv\cdot (\rv/t^{1/4}) -\hat\qv^4 }  \\
 &= \frac{2^{d/4-1/2} \pi^{1/2-d/2} }{ d\, \Gamma(1/2+d/4)\, r^{d-2}} \psi^{d/4-1/2} \Bigg[ \frac{d\, \Gamma(d/4-1/4)}{8\,\Gamma(d/4)}\, {}_1 F_3\left(\frac d 4 - \frac 1 2; \frac 1 2, \frac 1 2 + \frac d 4, \frac d 4; \psi \right) \\ 
&\qquad - \sqrt{\frac{\psi}{2}}\, {}_1 F_3\left(\frac d 4; \frac 3 2, \frac 1 2 + \frac d 4, 1+ \frac d 4; \psi \right)\Bigg], \qquad \psi \equiv \frac{r^4}{256 t},
\end{split}\label{eq_Cdyn_blk_crit}\eeq
where ${}_1 F_3$ is a generalized hypergeometric function \cite{olver_nist_2010} and, according to \cref{eq_time_resc}, the scaling variable $\psi$ is dimensionless.

\subsubsection{Film correlation function}
\label{sec_OP_correl_film}

In a thin film with periodic \bcs, translational invariance implies $\Ccal\pbc(\rvp, z,z',t) = \Ccal\pbc(\rvp,z-z',t)$ and the two-time correlation function is given by 
\beq 
\Ccal\pbc(\rvp, z- z', t) = \frac{1}{L} \sum_{n\in \mathbb{Z}} \int \frac{\d^{d-1}p}{(2\pi)^{d-1}}\, e^{\im \pv \cdot \rvp + \im k_n\ut{(p)} (z - z')} S(|\pv|, k_n\pbc, t),
\label{eq_Cdyn_film_pbc}
\eeq
where 
\beq k_n\pbc \equiv \frac{2\pi n}{L},\qquad n=0,\pm 1,\pm 2,\ldots\, 
\label{eq_eigenspec_pbc}\eeq
and 
\beq S(p,k,t) \equiv \frac{e^{-(p^2 + k^2 )(p^2 + k^2 + \tauLG ) t}}{p^2 + k^2 + \tauLG}.
\label{eq_S_correl_exp}\eeq
For Neumann and Dirichlet \bcs, respectively, one has
\beq 
\Ccal^{\text{(N)}\atop \text{(D)}}(\rvp, z, z', t) = \frac{1}{L} \sum_{n\in \begin{Bsmallmatrix} 
\mathbb{N}_0 \\
\mathbb{N}
\end{Bsmallmatrix}} \int \frac{\d^{d-1}p}{(2\pi)^{d-1}}\, e^{\im \pv \cdot \rvp}  
\begin{Bmatrix} 
\cos(k_n\Nbc z)\cos(k_n\Nbc z') \\
\sin(k_n\Dbc z)\sin(k_n\Dbc z')
\end{Bmatrix} S(|\pv|, k_n\ut{(N/D)}, t) ,
\label{eq_Cfour_film_NDbc}
\eeq 
with the eigenmodes 
\begin{subequations}
\begin{align}
k_n\Nbc &\equiv \frac{\pi n}{L},\qquad n=0, 1, 2,\ldots, \label{eq_eigenspec_Nbc} \\
k_n\Dbc &\equiv \frac{\pi n}{L},\qquad n= 1, 2,\ldots\,. \label{eq_eigenspec_Dbc}
\end{align}
\end{subequations}
Since (standard) Dirichlet \bcs are not flux conserving [see the discussion after \cref{eq_chempot}], we shall consider these \bcs only in the static limit given by setting $t=0$ in \cref{eq_Cfour_film_NDbc}.
By means of the Poisson resummation formula, the above correlation functions can be expressed in terms of $\Ccal_b$ as (see, e.g., Refs.\ \cite{gambassi_critical_2006,gross_dynamics_2019})
\begin{subequations}
\begin{align}
\Ccal\pbc(\rvp, z, t) &= \sum_{m=-\infty}^\infty \Ccal_b(\{ \rvp, z + m L \}, t),
\label{eq_C_film_pbc} \\
\Ccal\Nbc(\rvp, z,z',t) &= \Ccal\pbc(\rvp, z-z',t)\big|_{2L} + \Ccal\pbc(\rvp, z+z',t)\big|_{2L},
\label{eq_C_film_Neu} \\
\Ccal\Dbc(\rvp, z,z',t) &= \Ccal\pbc(\rvp, z-z',t)\big|_{2L} - \Ccal\pbc(\rvp, z+z',t)\big|_{2L},\label{eq_C_film_Dir}
\end{align}\label{eq_C_film}
\end{subequations}
\hspace{-0.15cm}where, as indicated by the subscript, $\Ccal\pbc$ on the r.h.s.\ of \cref{eq_C_film_Neu,eq_C_film_Dir} are obtained from \cref{eq_C_film_pbc} by the replacement $L\mapsto 2L$.
The bulk scaling behavior at short times, expressed in \cref{eq_Cdyn_blk_crit}, lends itself for being generalized to that of a film: 
\beq \Ccal(\rvp,z,z',t \ll \tauLG^{-2},\tauLG,L) = t^{-(d-1)/4} \widehat\Ccal(\rvp t^{-1/4}, z t^{-1/4}, z' t^{-1/4}, L t^{-1/4}) ,
\label{eq_Cdyn_film_shortT_scal}\eeq
with a scaling function $\widehat\Ccal$, the explicit expression of which is not needed in the following but can in principle be obtained from \cref{eq_Cdyn_blk_crit,eq_C_film}.

\subsubsection{Further properties and relations}

We close this section by collecting a number of relations useful for later developments in this study.
Owing to the linearity of the statistical average, one has 
\beq \bra \pd_\alpha^m \phi(\rv) \pd_\beta^{'n} \phi(\rv')\ket = \pd_\alpha^m \pd_\beta^{'n} \Ccal(\rv,\rv',t) , 
\label{}\eeq 
where $\alpha,\beta\in \{1,\ldots,d\}$ and $\pd_\alpha'\equiv \pd/\pd_{r'_\alpha}$.
Analogously, the Fourier and Poisson representations of the correlation function for the \bcs considered here [see \cref{eq_Cdyn_blk,eq_Cdyn_film_pbc,eq_Cfour_film_NDbc,eq_C_film}] render 
\beq \pd_z \Ccal(\rvp,z,z',t)\big|_{z=z'=0} = 0 = \pd_{z'} \Ccal(\rvp,z,z',t)\big|_{z=z'=0}
\label{eq_correl_pd_z0}\eeq 
in the bulk as well as in a film.
Moreover, for $\alpha\in \{1,\ldots, d-1\}$, one has
\beq \big\bra (\pd_\alpha \phi(\rvp,z,r))^2 \big\ket =  -\big\bra \phi(\rvp, z, t)  \pd_\alpha^2 \phi(\rvp, z, t) \big\ket =  -\pd_\alpha^2 \Ccal(\rvp, z,z,t) \big|_{\rvp=\bv0}, \label{eq_phi_lat_correlator}
\eeq 
where we used translational invariance in the lateral directions in order to obtain the last equation.

\section{Fluctuations of the CCF}
\label{sec_fluct_CCF}

In \emph{equilibrium}, the two-time correlation function of the instantaneous CCF follows from \cref{eq_CCF_Pfilm} and it is given by 
\beq \bra \Kcal(z,t) \Kcal(z',t')\ket = \bra \Pcal_f(z,t) \Pcal_f(z',t')\ket - \bra \Pcal_f(z)\ket \bra \Pcal_b(z')\ket - \bra \Pcal_b(z)\ket \bra \Pcal_f(z')\ket + \bra \Pcal_b(z,t)\Pcal_b(z',t')\ket,
\label{eq_Kvar}\eeq 
where we used the fact that film and bulk pressure are uncorrelated, 
\beq \bra \Pcal_f(z,t) \Pcal_b(z',t')\ket = \bra \Pcal_f(z,t) \ket\bra \Pcal_b(z',t')\ket,
\eeq 
since they emerge from physically separated parts of the system.
Accordingly, the fluctuations of the CCF (per area) given by 
\beq \Delta\Kcal(z,t) \equiv \Kcal(z,t)- \bra \Kcal(z) \ket
\label{eq_CCF_fluct_def}\eeq 
are correlated in equilibrium as 
\beq  \begin{split}
\bra \Delta\Kcal(z,t) \Delta\Kcal(z',t')\ket &= \bra \Kcal(z,t)\Kcal(z',t')\ket - \bra \Kcal(z)\ket \bra \Kcal(z')\ket \\
&= \bra \Pcal_f(z,t) \Pcal_f(z',t')\ket + \bra \Pcal_b(z,t)\Pcal_b(z',t')\ket  - \bra\Pcal_f(z)\ket\bra\Pcal_f(z')\ket - \bra\Pcal_b(z)\ket\bra\Pcal_b(z')\ket \\
&= \bra \Delta \Pcal_f(z,t) \Delta \Pcal_f(z',t')\ket + \bra \Delta \Pcal_b(z,t) \Delta \Pcal_b(z',t')\ket,
\end{split} \label{eq_fluct_CCF_correl}\eeq
with $\Delta\Pcal(z,t) \equiv \Pcal(z,t) - \bra\Pcal(z,t)\ket$ and where we used \cref{eq_Kvar} as well as 
\beq \bra\Kcal(z)\ket\bra\Kcal(z')\ket = \bra \Pcal_f(z)\ket\bra \Pcal_f(z')\ket - \bra\Pcal_f(z)\ket\bra\Pcal_b(z')\ket - \bra\Pcal_f(z')\ket\bra\Pcal_b(z)\ket + \bra \Pcal_b(z)\ket \bra \Pcal_b(z')\ket.
\eeq 
\Cref{eq_fluct_CCF_correl} reflects the fact that the variance of a sum of two uncorrelated random processes ($\Pcal_f$ and $-\Pcal_b$, see \cref{eq_CCF_Pfilm}) is the sum of the individual variances.
For $t-t'\to\infty$, $\Pcal_f(z,t)$ and $\Pcal_f(z',t')$ become uncorrelated in equilibrium, i.e., $\bra\Pcal_f(z,t)\Pcal_f(z',t')\ket\to \bra\Pcal_f(z)\ket \bra\Pcal_f(z')\ket$; this applies analogously to $\Pcal_b$.
Accordingly, one has 
\beq \bra \Delta\Kcal(z,t) \Delta\Kcal(z',t')\ket \overset{t-t'\to\infty}{=} 0.
\eeq 
Since we consider equilibrium dynamics, we can exploit time-translation invariance and henceforth set $t'=0$.

\subsection{Pressure correlation function}

\Cref{eq_fluct_CCF_correl} can be evaluated by using \cref{eq_CCF_Pfilm,eq_Pfilm,eq_Tzz_dyn} as well as the fact the four-point correlation function for a multivariate zero-mean Gaussian process $X_n \sim \phi$ or $X_n\sim \pd_\beta \phi$ is given by the following cumulant expansion:
\beq \bra X_i X_j X_k X_l\ket = \bra X_i X_j\ket\bra X_k X_l\ket + \bra X_i X_k\ket \bra X_j X_l\ket + \bra X_i X_l\ket \bra X_j X_k\ket.
\label{eq_Gauss_4pt}\eeq 
Furthermore, we use the fact that $\int_A \d^{d-1} r_\parallel \int_A \d^{d-1} r_\parallel'\, \Ccal(\rvp-\rvp') = A  \int_A \d^{d-1} r_\parallel\, \Ccal(\rvp)$, which follows upon applying the (volume preserving) coordinate transformation $(\rvp,\rvp')\mapsto (\rvp-\rvp',\rvp')$ and by using translational invariance in the lateral directions.
In order to illustrate the calculation of $\bra \Pcal_f(z,t)\Pcal_f(z',0)\ket$, we consider the term
\beq\begin{split} &\quad \frac{1}{4A^2} \int_A \d s_z(\rvp) \int_A \d s_z(\rvp') \big\bra [ \pd_z \phi(\rvp,z,t) ]^2 [\pd_{z'} \phi(\rvp',z',0) ]^2 \big\ket 
\\ &= \Pi + \frac{1}{2A^2} \int_A \d s_z(\rvp) \int_A \d s_z(\rvp') \big\bra \pd_z\phi(\rvp,z,t) \pd_{z'}\phi(\rvp',z',0) \big\ket^2 = \Pi + \frac{1}{2A} \int_A \d s_z(\Rv_\parallel) \left[ \pd_z\pd_{z'} \Ccal(\Rv_\parallel,z,z',t) \right]^2,
\end{split}\label{eq_Pf_correl_A}\eeq
where $\d s_z(\rv_\parallel)$ denotes an area element and 
\beq\begin{split} \Pi\, &\equiv \, \frac{1}{4A^2} \int_A \d s_z(\rvp)  \bra [\pd_z \phi(\rvp,z,t) ]^2\ket \int_A \d s_z(\rvp')\bra [\pd_{z'} \phi(\rvp',z',t) ]^2\ket \\
&= \frac{1}{4}  [\pd_z \pd_{z'} \Ccal\eq(\bv0, z,z')]_{z'\to z} [\pd_z \pd_{z'} \Ccal\eq(\bv0, z,z')]_{z\to z'} 
\end{split}\eeq 
represents a contribution to $\bra\Pcal_f(z)\ket \bra\Pcal_f(z')\ket$, resulting from the first term on the r.h.s.\ of \cref{eq_Gauss_4pt}.
In equilibrium, all these contributions are time-independent ($\Ccal\eq$ denotes the static equilibrium correlation function, see \cref{sec_OPcorrel}), and they are subtracted upon obtaining the CCF fluctuations according to \cref{eq_fluct_CCF_correl} [see also \cref{eq_CCF_eqvar} below]. 
Altogether, we obtain  
\begin{multline}
 \bra\Pcal_f(z,t)\Pcal_f(z',0)\ket = \bra\Pcal_f(z)\ket\bra\Pcal_f(z')\ket  + \frac{1}{A} \int_A \d^{d-1} r\, \Bigg\{ 
 %self-terms
 \frac{1}{2} (\pd_z \pd_{z'} \Ccal)^2 + \frac{1}{2} \sum_{\alpha,\beta=1}^{d-1} (-\pd_\alpha\pd_{\beta} \Ccal)^2 \\ 
 + \Ccal \sum_{\alpha,\beta=1}^d \pd_\alpha^2 \hat \pd_{\beta}^{2} \Ccal 
 + \Big(\sum_{\alpha=1}^d \pd_\alpha^2 \Ccal \Big) \Big(\sum_{\alpha=1}^d \hat\pd_\alpha^2 \Ccal \Big)   %cross-terms
 - \frac{1}{2} \sum_{\alpha=1}^{d-1} \left[ (\pd_z\pd_\alpha \Ccal)^2 + (\pd_{z'}\pd_\alpha \Ccal)^2 \right] \\
 - \Big[ (\pd_z \Ccal) \pd_z \sum_{\alpha=1}^d \hat \pd_\alpha^2 \Ccal + (\pd_{z'} \Ccal) \pd_{z'} \sum_{\alpha=1}^d \pd_\alpha^2 \Ccal \Big]
 +  \sum_{\alpha=1}^{d-1}  (\pd_\alpha\Ccal) \pd_\alpha \Big( \sum_{\beta=1}^d \pd_\beta^2 \Ccal + \sum_{\beta=1}^d \hat\pd_\beta^2 \Ccal \Big) 
 % tau terms
 +\frac{1}{2}\tauLG^2 \Ccal^2 \\ + \frac{1}{2}\tauLG \big[ (\pd_z \Ccal)^2  + (\pd_{z'} \Ccal)^2 \big] 
 - \tauLG \Big( \sum_{\alpha=1}^{d-1} \pd_\alpha \Ccal \Big)^2  - \tauLG  \Ccal \Big(\sum_{\alpha=1}^d \pd_\alpha^2 \Ccal + \sum_{\alpha=1}^d \hat\pd_\alpha^2 \Ccal \Big)
 \Bigg\},
\label{eq_Pf_correl}\end{multline}
where $\pd_\gamma \in \{\pd_{r_1},\ldots, \pd_{r_{d-1}}, \pd_{z}\}$ and we define $\hat\pd_\gamma \in \{\pd_{r_1},\ldots, \pd_{r_{d-1}}, \pd_{z'}\}$, such that $\hat\pd_\gamma = \pd_\gamma = \pd_{r_\gamma}$ for $\gamma=1,\ldots,d-1$ and $\hat\pd_{d}=\pd_{z'}$; for brevity, we suppress the arguments of $\Ccal(\rv,z,z',t)$ on the r.h.s.
The same expression results for $\bra\Pcal_b(z,t)\Pcal_b(z',0)\ket$, but with $\Ccal$ replaced by the bulk correlator $\Ccal_b$.
It is useful to note that the quantity $\bra \Delta \Pcal_f(z,t)\Delta \Pcal_f(z,0)\ket = \bra\Pcal_f(z,t)\Pcal_f(z',0)\ket - \bra\Pcal_f(z)\ket\bra\Pcal_f(z')\ket$ has the length dimension of $A^{-1} L^{-(d+1)}$, which can be readily inferred from \cref{eq_Pf_correl,eq_Cdyn_film_pbc,eq_Cfour_film_NDbc}.

We now specialize \cref{eq_Pf_correl} to various \bcs in thin films and make use of the available spatial symmetries.
For Neumann and Dirichlet \bcs we will obtain expressions in terms of the correlation function for periodic \bcs. 
The film pressure variances are evaluated further in \cref{sec_stat_press,sec_dyn_press}.

\subsubsection{Bulk and periodic BCs}
\label{sec_presscorr_bulkper}

In the following, we jointly analyze the pressure both for the bulk system and for thin films with periodic \bcs in the transverse direction.
We note that translational invariance [see \cref{eq_Cdyn_film_pbc}] implies 
\beq 
\big\bra [\pd_z \phi(\rvp, z, t)]^2 \big\ket = -\big\bra \phi(\rvp, z, t) \pd_z^2 \phi(\rvp, z, t) \big\ket = - \pd_z^2 \Ccal\pbc(\rvp=\bv0, z,t)\big|_{z=0},
\label{eq_phi_z_correlator_pbc}
\eeq
with the analogous expression applying in the bulk.
Upon using \cref{eq_phi_lat_correlator,eq_phi_z_correlator_pbc}, the mean equilibrium CCF for a film, with one surface at $z=0$ and with periodic \bcs, follows from \cref{eq_CCF_Tzz} \cite{gross_dynamics_2019}:
\beq \bra \Kcal\pbc \ket = -\frac{3}{2}\pd_z^2 \Ccal\pbc\eq(\rvp=\bv0, z)\big|_{z=0} - \frac{1}{2} \sum_{\alpha=1}^{d-1} \pd_\alpha^2 \Ccal\pbc\eq(\rvp,0)\big|_{\rvp=\bv0}.
\eeq 
The film pressure correlator in \cref{eq_Pf_correl} reduces to 
\begin{multline}
 \bra\Pcal_f\pbc(z,t)\Pcal_f\pbc(z',0)\ket = \bra\Pcal_f\pbc\ket^2  + \frac{1}{A} \int_A \d^{d-1} r\, \Big\{ 
 %self-terms
 \frac{1}{2} (-\pd_z^2  \Ccal\pbc)^2 + \frac{1}{2} \sum_{\alpha,\beta=1}^{d-1} (-\pd_\alpha\pd_{\beta} \Ccal\pbc)^2 \\ + \Ccal\pbc \sum_{\alpha,\beta=1}^d \pd_\alpha^2 \pd_{\beta}^2 \Ccal\pbc  + \Big(\sum_{\alpha=1}^d \pd_\alpha^2 \Ccal\pbc \Big)^2 
 %cross-terms
 - \sum_{\alpha=1}^{d-1} (\pd_z\pd_\alpha \Ccal\pbc)^2  -  2  (\pd_z \Ccal\pbc) \pd_z \sum_{\alpha=1}^d \pd_\alpha^2 \Ccal\pbc  \\
 + 2\sum_{\alpha=1}^{d-1}  (\pd_\alpha\Ccal\pbc) \pd_\alpha \sum_{\beta=1}^d \pd_\beta^2 \Ccal\pbc 
 % \tau terms
 +\frac{1}{2}\tauLG^2 (\Ccal\pbc)^2  + \tauLG \left( \pd_z \Ccal\pbc \right)^2 
 - \tauLG \Big( \sum_{\alpha=1}^{d-1} \pd_\alpha \Ccal\pbc\Big)^2  - 2\tauLG \Ccal\pbc \sum_{\alpha=1}^d \pd_\alpha^2 \Ccal\pbc  \Big\},
\label{eq_Pf_correl_pbc}\end{multline}
where we again suppressed the arguments of $\Ccal(\rv,z-z',t)$ on the r.h.s.
In the special case $z=z'=0$, \cref{eq_Pf_correl_pbc} can be simplified using \cref{eq_correl_pd_z0}:
\begin{multline}
 \bra\Pcal_f\pbc(z=0,t)\Pcal_f\pbc(z'=0,0)\ket = \bra\Pcal_f\pbc\ket^2  + \frac{1}{A} \int_A \d^{d-1} r\, \Big\{ 
 %self-terms
 \frac{1}{2} (-\pd_z^2  \Ccal\pbc)^2 \\ + \frac{1}{2} \sum_{\alpha,\beta=1}^{d-1} \left(-\pd_\alpha\pd_{\beta} \Ccal\pbc\right)^2 + \Ccal\pbc \sum_{\alpha,\beta=1}^d \pd_\alpha^2 \pd_{\beta}^2 \Ccal\pbc  + \Big(\sum_{\alpha=1}^d \pd_\alpha^2 \Ccal\pbc\Big)^2 
 %cross-terms
  + 2\sum_{\alpha=1}^{d-1}  (\pd_\alpha\Ccal\pbc) \pd_\alpha \sum_{\beta=1}^d \pd_\beta^2 \Ccal\pbc \\
 % \tau terms
 +\frac{1}{2}\tauLG^2 (\Ccal\pbc)^2  
  - \tauLG \Big( \sum_{\alpha=1}^{d-1} \pd_\alpha \Ccal\pbc\Big)^2  - 2\tauLG \Ccal\pbc \sum_{\alpha=1}^d \pd_\alpha^2 \Ccal\pbc
 \Big\}_{z=z'=0}.
\label{eq_Pf_correl_pbc_z0}\end{multline}
For bulk pressure correlations $\bra\Pcal_b(z,t)\Pcal_b(z',0)\ket$, the same expressions as in \cref{eq_Pf_correl_pbc,eq_Pf_correl_pbc_z0} apply, but with $\Ccal\pbc$ replaced by the bulk correlation function $\Ccal_b$.

\subsubsection{Neumann \bcs}
\label{sec_presscorr_Neu}

The pressure correlation function for a film with Neumann \bcs is given by \cref{eq_Pf_correl} with $\Ccal$ replaced by $\Ccal\Nbc$ [see \cref{eq_C_film_Neu}].  
In the special case $z=z'=0$, the resulting expression can be simplified by noting that \cref{eq_C_film_Neu} implies
\begin{subequations}
\begin{align}
\pd_{z'} \Ccal\Nbc(\rvp, z,z',t)\big|_{z=z'=0} &= 0, \\
\pd_z \Ccal\Nbc(\rvp,z,z',t)\big|_{z=z'=0} &= 2\pd_z\Ccal\pbc(\rvp,z=0,t)\big|_{2L} = 0, \\
\pd_\alpha^n \Ccal\Nbc(\rvp,z,z',t)\big|_{z=z'=0} &= 2 \pd_\alpha^n \Ccal\pbc(\rvp,0,t)\big|_{2L},\qquad \alpha\in\{1,\ldots,d-1\}, \\
\pd_z^2 \Ccal\Nbc(\rvp, z, z', t)\big|_{z=z'=0} &= \pd_{z'}^2 \Ccal\Nbc(\rvp,z,z',t)\big|_{z=z'=0} = 2\pd_z^2 \Ccal\pbc(\rvp,z=0,t)\big|_{2L}.
\end{align}\label{eq_correl_pd_z0_Nbc}
\end{subequations}
Using these relations in \cref{eq_Pf_correl} renders
\begin{multline}
 \bra\Pcal_f\Nbc(z=0,t)\Pcal_f\Nbc(z'=0,0)\ket = \bra\Pcal_f\Nbc\ket^2  + \frac{1}{A} \int_A \d^{d-1} r\, \Bigg\{ 
 %self-terms
 2 \sum_{\alpha,\beta=1}^{d-1} (-\pd_\alpha\pd_{\beta} \Ccal\pbc)^2 \\ + 4\Ccal\pbc \sum_{\alpha,\beta=1}^d \pd_\alpha^2 \pd_{\beta}^{2} \Ccal\pbc + 4\Big(\sum_{\alpha=1}^d \pd_\alpha^2 \Ccal\pbc\Big)^2 
 %cross-terms
 + 8 \sum_{\alpha=1}^{d-1}  (\pd_\alpha\Ccal\pbc) \pd_\alpha \sum_{\beta=1}^d \pd_\beta^2 \Ccal\pbc \\
 % \tau terms
 + 2\tauLG^2 (\Ccal\pbc)^2  
 - 4\tauLG \Big( \sum_{\alpha=1}^{d-1} \pd_\alpha \Ccal\pbc\Big)^2 - 8\tauLG \Ccal\pbc \sum_{\alpha=1}^d \pd_\alpha^2 \Ccal\pbc \Bigg\}_{z=z'=0,\atop L\mapsto 2L},
\label{eq_Pf_correl_Nbc}\end{multline}
where, as indicated on the r.h.s., the correlation function $\Ccal\pbc$ for periodic \bcs has to be evaluated for a film thickness of $2L$.

\subsubsection{Dirichlet \bcs}
We begin by noting that \cref{eq_C_film_Dir} implies 
\begin{subequations}
\begin{align}
\Ccal\Dbc(\rvp,z=0,z'=0,t) &= 0,\\
\pd_{z'} \Ccal\Dbc(\rvp, z,z',t)\big|_{z=z'=0} &= -2\pd_z\Ccal\pbc(\rvp,z=0,t)\big|_{2L} = 0, \\
\pd_z^n \Ccal\Dbc(\rvp,z,z',t)\big|_{z=z'=0} &= 0, \\
\pd_\alpha^n \Ccal\Dbc(\rvp,z,z',t)\big|_{z=z'=0} &= 0 ,\\
\pd_{z'}^2 \Ccal\Dbc(\rvp,z,z',t)\big|_{z=z'=0} &= 0, \\
\pd_z \pd_z \Ccal\Dbc(\rvp,z,z',t)\big|_{z=z'=0} &= -2 \pd_z^2 \Ccal\pbc(\rvp,z=0, t)\big|_{2L}.
\end{align}\label{eq_C_film_Dir_relations}
\end{subequations}
Using these relations to evaluate the film pressure correlator in \cref{eq_Pf_correl} for $z=z'=0$ renders
\begin{equation}
 \bra\Pcal_f\Dbc(z=0,t)\Pcal_f\Dbc(z'=0,0)\ket = \bra\Pcal_f\Dbc\ket^2 + \frac{1}{A} \int_A \d^{d-1} r\, \Big\{ 
 %self-terms
 2 [-\pd_z^2  \Ccal\pbc(\rv,z,t)]^2  
 %cross-terms
\Big\}_{z=0, \atop L\mapsto 2L}.
\label{eq_Pf_correl_Dbc}\end{equation}

\subsection{Equilibrium variance}
\label{sec_stat_press}

The equilibrium variance of the CCF at a fixed boundary (located at $z=z'=0$) follows by setting $t=t'=0$ in \cref{eq_fluct_CCF_correl}, resulting in 
\beq  \begin{split} \bra \Delta\Kcal^2\ket = \bra \Kcal^2\ket - \bra \Kcal\ket^2 &= \bra \Pcal_f^2\ket - \bra\Pcal_f\ket^2 + \bra \Pcal_b^2\ket - \bra\Pcal_b\ket^2 = \left\bra (\Delta \Pcal_f)^2 \right\ket + \left\bra (\Delta \Pcal_b)^2 \right\ket .
\end{split} 
\label{eq_CCF_eqvar}\eeq
This expression is evaluated in the following by replacing in \cref{eq_Pf_correl} $\Ccal$ by the expression of $\Ccal\eq$ for the respective \bcs.
Below we discuss the final results at bulk criticality ($\tauLG=0$).

\subsubsection{Bulk system}

The static equilibrium variance of the bulk pressure $\bra \Delta\Pcal_b^2\ket  \equiv  \bra \Pcal_b^2\ket - \bra\Pcal_b\ket^2$ is obtained by inserting the equilibrium bulk correlation function $\Ccal\eqBlk$ [\cref{eq_Ceq_blk}] for $\Ccal$ into \cref{eq_Pf_correl}.
In order to simplify this calculation, we note that, due to the translational invariance of $\Ccal_b$, $\bra \Delta\Pcal_b^2\ket$ takes the same form as the variance for periodic \bcs reported in \cref{eq_Pf_correl_pbc_z0} (with $\Ccal\pbc$ replaced by $\Ccal_b$).
We furthermore recall the Schwinger-Dyson relation (see, e.g., Ref.\ \cite{gross_dynamics_2019})
\beq \left\bra \phi(\rv) \frac{\delta \Fcal}{\delta \phi(\rv')} \right\ket = \delta(\rv-\rv') ,
\eeq 
which implies the following identity for the equilibrium bulk correlation function:
\beq -\nabla^2 \Ccal\eqBlk(\rv) + \tau \Ccal\eqBlk(\rv) = \delta(\rv).
\label{eq_Cblk_Ward}\eeq 
Using this relation in \cref{eq_Pf_correl_pbc_z0} renders
\begin{multline}
  \bra \Delta\Pcal_b^2\ket \equiv  \bra \Pcal_b^2\ket - \bra\Pcal_b\ket^2 
 =  \frac{1}{A} \Big\{ - \bv{n}\cdot \big[\Ccal\eqBlk \nabla_\parallel\delta(\rvp) + \delta(\rvp)\nabla_\parallel \Ccal\eqBlk\big]_{\partial A}\delta(z) - \Ccal\eqBlk(\bv0_\parallel,z) \pd_z^2\delta(z) - \delta(z) \pd_z^2 \Ccal\eqBlk(\bv0_\parallel,z) \Big\}_{z=0} \\
 + \frac{1}{A} \int_A \d^{d-1} r\, \Big\{ 
 \frac{1}{2} [-\pd_z^2  \Ccal\eqBlk]^2 + \frac{1}{2} \sum_{\alpha,\alpha'}^{d-1} [-\pd_\alpha\pd_{\alpha'} \Ccal\eqBlk]^2 
 +\frac{1}{2}\tauLG^2 \Ccal\eqBlk^2  + \tauLG \Big[ \nabla_\parallel \Ccal\eqBlk\Big]^2 \Big\}_{z=0},
\label{eq_Pb_correl}\end{multline}
where $\nv$ denotes the unit normal of the $(d-1)$-dimensional surface plane.

Upon inserting $\Ccal\eqBlk$ [\cref{eq_Ceq_blk}] into \cref{eq_Pb_correl}, one finds that the resulting integral over $A$ generally converges at its upper limit of large distances $r$ (where $r$ refers to the spherical coordinate), allowing one to set $A\to \infty$ for its evaluation.
However, the integral does not converge at the lower limit $r\to 0$, requiring a regularization via a small-distance cutoff $\varepsilon$.
Similarly, a regularization is also necessary for the terms in the first line of \cref{eq_Pb_correl}. To this end, we evaluate their Fourier space representation by taking into account a wavenumber cutoff $q\st{max}\simeq 1/\cutoff$ and, accordingly, regularize the $\delta$ function as \cite{bartolo_fluctuations_2002} $\delta(z) = \int_{-1/\cutoff}^{1/\cutoff} \frac{\d q}{2\pi} \exp(\im q z)$, which results in 
\beq \delta(0)\big|_\reg = \frac{1}{\pi \cutoff},\qquad \pd_z^2 \delta(z=0)\big|_\reg = - \frac{1}{3\pi \cutoff^3}.
\eeq 
Analogously, based on the Fourier representation in \cref{eq_Cdyn_blk}, we obtain the regularized form of the bulk correlator (at $\tauLG=0$)
\begin{subequations}\begin{align}
\Ccal\eqBlk(\bvnp, 0)\big|_\reg &= \frac{2 }{(4\pi)^{d/2}(d-2) \Gamma(d/2)} \frac{1}{\cutoff^{d-2}},\\
\pd_z^2 \Ccal\eqBlk(\bvnp, z=0)\big|_\reg &= -\frac{1}{(4\pi)^{d/2} d \Gamma((d+1)/2)} \frac{1}{\cutoff^d}.
\end{align}\end{subequations} 
The term $-\nv\cdot[\ldots]_{\pd A}\delta(z)$ in \cref{eq_Pb_correl} vanishes because the boundary $\pd A$ of the surface is assumed to be located basically at infinity, such that $r_{\parallel,\alpha} \neq 0$.

Altogether, the variance of the bulk pressure at bulk criticality ($\tauLG=0$) turns out to be 
\begin{multline} \bra \Delta\Pcal_b^2\ket \equiv  \bra \Pcal_b^2\ket - \bra\Pcal_b\ket^2 
= \left[ \frac{d(d-1) \Gamma(d/2)^2}{4 \pi^{(d+1)/2} (d+1) \Gamma((d-1)/2)} + \frac{(d^2+3d-6)}{3\times 2^d\pi^{d/2+1} d (d-2) \Gamma(1+d/2) } \right] \frac{1}{A \cutoff^{d+1}}.
\label{eq_var_CCF_bulk_pbc}\end{multline}
This expression is singular at $d=2$ and positive for all other integer dimensions $d$.

\subsubsection{Periodic BCs}

According to \cref{eq_C_film_pbc}, the static OP correlator for a thin film with periodic \bcs follows as
\beq\begin{split} 
\Ccal\eq\pbc(r_\parallel, z) &= \sum_{m=-\infty}^\infty \Ccal\eqBlk(\{r_\parallel, z+ mL\}) 
= \frac{\Gamma(d/2-1)}{4\pi^{d/2}} \sum_{m=-\infty}^\infty \frac{1}{\left|(z+m L)^2 + r_\parallel^2 \right|^{(d-2)/2}} \\
&= \frac{\Gamma(d/2-1)}{4\pi^{d/2} L^{d-2}} \sum_{m=-\infty}^\infty \frac{1}{\left|(\hat z + m )^2 + \hat r_\parallel^2 \right|^{(d-2)/2}} \equiv L^{2-d} \hat \Ccal\eq\pbc(\hat r_\parallel, \hat z),
\end{split}\label{eq_Ceq_film}\eeq 
where $\Ccal\eqBlk(r)$ denotes the bulk correlation function [\cref{eq_Ceq_blk}]. In the last line we introduced the dimensionless coordinates $\hat z\equiv z/L$ and $\hat r_\parallel \equiv r_\parallel/L$ and defined the scaling function $\hat \Ccal\eq\pbc$.
We note that in \cref{eq_Ceq_film} the bulk correlation function corresponds  to the term with $m=0$.

Using \cref{eq_C_film_pbc} in \cref{eq_Cblk_Ward} yields the following identity for the film correlation function:
\beq -\nabla^2 \Ccal\eq\pbc(\rv_\parallel,z) + \tauLG\Ccal\eq\pbc(\rv_\parallel,z)  = \delta(\rv_\parallel) \sum_{m=-\infty}^\infty \delta(z+ mL).
\label{eq_Cfilm_Ward}\eeq 
Using this relation in \cref{eq_Pf_correl_pbc} with $z=z'=0$ renders for $\bra (\Delta\Pcal_f\pbc)^2\ket$ the same form as in \cref{eq_Pb_correl}, with $\Ccal\eqBlk$ replaced by $\Ccal\pbc\eq$.
Furthermore, upon inserting \cref{eq_Ceq_film} into the resulting expression for $\bra (\Delta\Pcal_f\pbc)^2\ket$ and using the relation $\pd_\beta \Ccal\eq(\rvp, z) = L^{-1} \pd_{\hat \beta} \hat\Ccal\eq(\hat r_\parallel, \hat z)$, the integral over the transverse area $A$ converges for $m\neq 0$ and results in expressions of the form $\sum_{m_1=1}^\infty \sum_{m_2=1}^\infty (m_1+m_2)^{-d-1}$, which can be evaluated in terms of the Riemann zeta function $\zeta$. For $m_1=0$ or $m_2=0$, the integral over $A$ does not converge at its lower limit ($r_\parallel\to 0$) and is thus evaluated using a small-distance cutoff $\cutoff$ for the radial coordinate $r_\parallel$. 
The integral over $A$ generally converges at its upper (large distance) limit. 
Altogether, we obtain the following film pressure variance for periodic \bcs with $\tauLG=0$:
\begin{multline} \bra (\Delta\Pcal_f\pbc)^2\ket = \bra \Pcal_f^2\ket - \bra\Pcal_f\ket^2  =  \frac{1}{A L^{d+1}} \Bigg[ d(d-1) \pi^{-d/2} \Gamma(d/2) \zeta(d) - \frac{d\, \Gamma(d/2)^2 \zeta(d)}{\pi^{(d+1)/2} \Gamma((d-1)/2)} \frac{L}{\cutoff}  \Bigg] +  \bra\Delta\Pcal_b^2\ket.
\label{eq_var_CCF_film_pbc}\end{multline} 
The power laws for $L$ and $\cutoff$ emerge by using the scaling form for $\Ccal\eqBlk$ given in \cref{eq_Ceq_film}. 
It turns out that $\bra (\Delta\Pcal_f\pbc)^2\ket A L^{d+1}$, which is only a function of $d$ and the dimensionless parameter $L/\varepsilon$, remains positive for all $d>2$ and all $L/\varepsilon>0$. In the relevant regime $L/\varepsilon\gg 1$, the variance $\bra (\Delta\Pcal_f\pbc)^2\ket$ is, in fact, dominated by $\bra\Delta\Pcal_b^2\ket \propto 1/(A\varepsilon^{d+1})$ [see \cref{eq_var_CCF_bulk_pbc}].

\subsubsection{Neumann \bcs}

\Cref{eq_Pf_correl_Nbc} renders, after some algebra analogous to the one leading to \cref{eq_var_CCF_film_pbc}, the static variance 
\begin{multline} \bra (\Delta\Pcal_f\Nbc)^2\ket = \bra (\Pcal_f\Nbc)^2\ket - \bra\Pcal_f\Nbc\ket^2  =  \frac{1}{A L^{d+1}} \Bigg[   \frac{d(d-1) \Gamma(d/2) \zeta(d)}{(4\pi)^{d/2}}  - \frac{\Gamma(d/2)^2\zeta(d)}{2^{d-2} \pi^{(d+1)/2} \Gamma((d-1)/2)} \frac{L}{\cutoff}    \Bigg] + \\
 \frac{1}{A \cutoff^{d+1}} \left[ \frac{(d^2-d-1) \Gamma(d/2)^2}{\pi^{(d+1)/2} (d+1) \Gamma((d-1)/2)} + \frac{4(d^2+3d-6)}{3\times 2^d\pi^{d/2+1} d (d-2) \Gamma(1+d/2) } \right] ,
\label{eq_var_CCF_film_Nbc}
\end{multline}
where $\varepsilon$ denotes a small length scale required for regularizing the integral over the surface area. The term $\propto 1/(A \varepsilon^{d+1})$ scales like the bulk pressure variance [see \cref{eq_var_CCF_bulk_pbc}], but its detailed form differs from the one in \cref{eq_var_CCF_film_pbc} due to the different structures of \cref{eq_Pf_correl_Nbc,eq_Pf_correl_pbc_z0}.

\subsubsection{Dirichlet \bcs}

Upon introducing, as above, a small-distance cutoff $\varepsilon$ at the lower limit of the areal integral in \cref{eq_Pf_correl_Dbc}, the static pressure variance follows as \footnote{The singular terms in the first line in \cref{eq_Pb_correl} vanish for Dirichlet \bcs in a cutoff-regularized field theory [see \cref{eq_C_film_Dir_relations}] and we thus neglected them here.}
\begin{multline} \bra (\Delta\Pcal_f\Dbc)^2\ket = \bra (\Pcal_f\Dbc)^2\ket - \bra\Pcal_f\Dbc\ket^2  =  \frac{1}{A L^{d+1}} \Bigg[  \frac{ d(d-1)\Gamma(d/2) \zeta(d) }{(4\pi)^{d/2} }  - \frac{(d-1)\Gamma(d/2)^2\zeta(d)}{ 2^{d-2} \pi^{(d+1)/2} \Gamma((d-1)/2) } \frac{L}{\cutoff}  \Bigg]  + \\
\frac{1}{A \cutoff^{d+1}}  \frac{\Gamma(d/2)^2}{2 \pi^{(d+1)/2} (d+1) \Gamma((d-1)/2)}. 
\label{eq_var_CCF_film_Dbc}\end{multline} 
The universal subdominant term (first term in the square brackets) coincides with Eq.~(14) in Ref.\ \cite{bartolo_fluctuations_2002}, as does the dominant (bulk-like) scaling behavior for small $\varepsilon$ [given by the last term in \cref{eq_var_CCF_film_Dbc} and by Eq.~(13) in Ref.\ \cite{bartolo_fluctuations_2002}].

\subsubsection{Discussion}

According to \cref{eq_CCF_eqvar}, the variance of the CCF (per area $A$ and thermal energy $k_B T$) is given by the sum of the bulk and film pressure variances.
The film and bulk pressure variances determined above [see \cref{eq_var_CCF_bulk_pbc,eq_var_CCF_film_pbc,eq_var_CCF_film_Nbc,eq_var_CCF_film_Dbc}] pertain to a thin film geometry ($A=L_\parallel^{d-1}\to\infty$) and thus represent the leading terms in an expansion in terms of $1/A$ (for $\varepsilon$ nonzero).
We thus expect the above results to represent a reasonable estimate for systems with a sufficiently large aspect ratio $L_\parallel/L$, such that the lateral OP modes approximately form a continuum spectrum.
Thus, in the limit of small $\cutoff\to 0$, the static equilibrium variance of the CCF turns out to be dominated by bulk-like contributions, which induce the scaling behavior 
\beq \bra (\Delta\Kcal)^2\ket^{1/2}  \propto \frac{1}{A^{1/2} \cutoff^{(d+1)/2}}
\label{eq_rms_CCF}\eeq 
where the proportionality constant is of $\Ocal(1)$ [see \cref{eq_var_CCF_film_pbc,eq_var_CCF_film_Nbc,eq_var_CCF_film_Dbc}].
We emphasize that \cref{eq_rms_CCF} quantifies the typical fluctuations of the Casimir \emph{pressure} $\Kcal$. Those of the actual Casimir \emph{force} acting on a surface of area $A$ are characterized by $A \bra (\Delta\Kcal)^2\ket^{1/2}$ and thus exhibit the expected thermodynamic scaling behavior \cite{dean_fluctuation_2013}. 
The essential scaling behavior expressed in \cref{eq_rms_CCF} has been confirmed by Monte Carlo simulations of various lattice models (see Refs.\ \cite{dantchev_critical_2004} and \footnote{In Ref.\ \cite{dantchev_critical_2004}, a different definition of the variance has been used, leading to slightly different geometric factors compared to \cref{eq_rms_CCF} and Ref.\ \cite{bartolo_fluctuations_2002}.}).
Compared to the mean values of the CCF \cite{krech_free_1992},
\begin{subequations}
\begin{align}
\bra \Kcal\pbc\ket &= L^{-d}  \pi^{-d/2} \Gamma(d/2) (1-d) \zeta(d) \overset{d=3}{\simeq}  -\frac{0.38}{L^3}, \label{eq_mean_CCF_film_pbc} \\
\bra \Kcal\Nbc\ket &= 2^{-d} \bra \Kcal\pbc\ket, \label{eq_mean_CCF_film_Nbc} \\
\bra \Kcal\Dbc\ket &= 2^{-d} \bra \Kcal\pbc\ket, \label{eq_mean_CCF_film_Dbc}
\end{align}\label{eq_mean_CCF}
\end{subequations}
\hspace{-0.11cm}the dispersion $\bra (\Delta\Kcal)^2\ket^{1/2} $ of the CCF is orders of magnitude larger and non-universal, as noted previously in Ref.\ \cite{bartolo_fluctuations_2002}.
In order to address this issue, it has been argued in Ref.\ \cite{bartolo_fluctuations_2002} that the static variance is unobservable because any measurement device experiences a force averaged over a \emph{finite} time interval. The variance of the average force $\bar\Kcal$ is thus reduced by a factor $N=t\st{res}/t\st{corr}$, where $t\st{res}$ is the temporal resolution of the measurement device and $t\st{corr}$ is the correlation time of short-wavelength fluctuations (which provide the dominant contribution to the variance). For typical experimental setups, $N$ is estimated to be of $\Ocal(10^4)$, which reduces the dispersion $\bra \Delta \bar\Kcal^2\ket^{1/2}= N^{-1/2} \bra\Delta \Kcal^2\ket^{1/2}$ to a value comparable to the mean force $\bra\Kcal\ket$ \cite{bartolo_fluctuations_2002}.

In the remaining part of this study, we shall analyze temporal pressure correlations as well as experimentally observable quantities influenced by the fluctuations of the CCF. We find that in these cases the cutoff dependence is mitigated or even disappears, without the need to invoke the temporal resolution of the measurement device and to formulate specific assumptions about it.

\subsection{Dynamic pressure correlations}
\label{sec_dyn_press}

We first focus on a film with periodic \bcs.
The associated two-time correlation function of the CCF at a fixed surface ($z=z'=0$) results from \cref{eq_fluct_CCF_correl} as 
\beq \bra \Delta\Kcal(t) \Delta\Kcal(0)\ket = \bra \Delta\Pcal_b(t)\Delta\Pcal_b(0)\ket + \bra \Delta\Pcal_f\pbc(t) \Delta\Pcal_f\pbc(0)\ket .
\label{eq_CCF_correl_dyn}\eeq 
In the following, we evaluate the dynamic bulk and film pressure correlations based on the formalism developed in \cref{sec_presscorr_bulkper,sec_presscorr_Neu}.

\subsubsection{Bulk pressure correlations}
\label{sec_dyn_press_bulk}

In the \emph{short-time limit} $t\ll \tauLG^{-2}$ (which includes the critical case $\tauLG=0$), terms proportional to $\tauLG$ or $\tauLG^2$ in \cref{eq_Pf_correl} are subdominant and can be asymptotically neglected \footnote{This can be shown by inserting the critical bulk correlator $\Ccal_b$ given in \cref{eq_Cdyn_blk_crit} into \cref{eq_Pf_correl_pbc_z0}.}.
Accordingly, in this regime, the dynamic bulk pressure correlations at a surface located at $z=0$ are obtained by inserting \cref{eq_Cdyn_blk_crit} into \cref{eq_Pf_correl_pbc_z0}:
\beq\begin{split} 
\bra \Delta\Pcal_b(t)\Delta\Pcal_b(0)\ket\big|_{t\ll \tauLG^{-2}} &= \bra\Pcal_b(t)\Pcal_b(0)\ket - \bra\Pcal_b\ket^2 \\
&= \frac{H_d}{A} t^{-(d+1)/4} ,\quad \text{with}\quad H_d\equiv \Omega_{d-1} \int_0^\infty \d\psi\, F_d(\psi),
\end{split}\label{eq_Pb_dyn_shortT}\eeq 
where
\beq \Omega_d = \frac{2\pi^{d/2}}{\Gamma(d/2)}
\label{eq_unitball}\eeq 
is the surface area of the $d$-dimensional unit sphere and $F_d(\psi)$ is a time-independent function which is constant for $\psi\to 0$ and $d> 2$, while it vanishes exponentially for $\psi\to\infty$. Its explicit form is rather lengthy and thus it is not reported here.
A numerical evaluation of $H_d$, which is finite in spatial dimensions $d>2$, gives $H_3\simeq 5.299\times 10^{-3}$ and $H_4\simeq 5.924\times 10^{-4}$.

Conversely, at \emph{long times} $t\gg \tauLG^{-2}$, using the scaling form given in \cref{eq_Cdyn_larget} for \cref{eq_Pf_correl_pbc_z0}, the bulk pressure correlations are dominated by the term $(1/2)\tauLG^2 \Ccal(\rvp,0,t)^2$. Evaluating the remaining integral over $\rvp$ renders
\beq \bra \Delta\Pcal_b(t)\Delta\Pcal_b(0)\ket\big|_{t\gg\tauLG^{-2}} = \frac{1}{2^{3(d+1)/2} \pi^{(d+1)/2} A} (\tauLG t)^{-(d+1)/2} .
\label{eq_Pb_dyn_lateT}\eeq 

\subsubsection{Periodic \bcs}
\label{sec_dyncor_per}

In order to determine the dynamic correlations of the film pressure at a fixed surface ($z=0$), we insert \cref{eq_Cdyn_film_pbc} into \cref{eq_Pf_correl_pbc_z0}.
A typical term is given by, e.g.,
\beq\begin{split} &\frac{1}{A} \int \d^{d-1}r\, [\pd_\alpha^2 \Ccal\pbc(\rv,z-z',t)][\pd_\beta^2 \Ccal\pbc(\rv,z-z',t)]_{z=z'} \\
& = \frac{1}{A L^2} \int \d^{d-1}r \sum_{m,n=-\infty}^\infty \int \frac{\d^{d-1} p}{(2\pi)^{d-1}} \int \frac{\d^{d-1} \tilde p}{(2\pi)^{d-1}}   p_\alpha^2 \tilde p_\beta^2 e^{\im(\pv+ \tilde \pv)\cdot \rv + \im (z-z') k_m - \im (z-z')k_n} S(p,k_m,t) S(\tilde p,k_n,t)\big|_{z-z'=0} \\
& = \frac{1}{A L^2} \sum_{m,n=-\infty}^\infty \int \frac{\d^{d-1} p}{(2\pi)^{d-1}}  p_\alpha^2 p_\beta^2 S(p,k_m,t) S(p,k_n,t),
\end{split}\label{eq_Pf_correl_sample}\eeq 
where $S(p,k,t)$ is reported in \cref{eq_S_correl_exp}.
In summary, the dynamic film pressure correlation at a fixed surface is given by
\begin{multline} 
\bra \Delta\Pcal_f\pbc(t)\Delta\Pcal_f\pbc(0)\ket = \bra\Pcal_f\pbc(t)\Pcal_f\pbc(0)\ket - \bra\Pcal_f\pbc\ket^2 \\
= \frac{K_{d-1}}{2 A L^2} \sum_{m,n=-\infty}^\infty  \int_0^\infty \d p\, S(p,k_m,t) S(p,k_n,t) p^{d-2} \Big[ k_m^2 k_n^2 + 2(p^2+ k_n^2)(k_m^2+ k_n^2) + p^4  
+2\tauLG(2k_n^2 + p^2) + \tauLG^2 \Big],
\label{eq_Pf_dyn_correl_pbc}\end{multline}
with $K_d \equiv \Omega_d/(2\pi)^{d}$ and $\Omega_d$ reported in \cref{eq_unitball}.

The integrand $\Ical(p,k_m,k_n)$ in \cref{eq_Pf_dyn_correl_pbc} has the symmetry $\Ical(p,k_m,k_n) = \Ical(p,-k_m, -k_n)$. While it is not symmetric upon exchanging $k_m$ and $k_n$, it can be written in such a form by using the fact that $m$ and $n$ are summed over. (In the following discussion, $k_m$ and $k_n$ refer to the second and third argument of $\Ical$, respectively.)
For small $p$ and $\tauLG=0$, the integrand behaves as $\Ical \propto p^{d-2}$ if $k_m\neq 0$ as well as if $k_m=k_n=0$, while $\Ical \propto p^{d-4}$ if $k_m=0$, $k_n\neq 0$; for $\tauLG\neq 0$ one has $\Ical\propto p^{d-2}$ at small $p$.
Accordingly, for $\tauLG=0$ and $d\leq 3$, the integral in \cref{eq_Pf_dyn_correl_pbc} is infrared divergent. This divergence is cut off by a finite lateral extent of the confining surfaces. 
For $t=0$, the integrand diverges as $\Ical \propto p^{d-2}$ for large $p$, which reflects the divergence of the static equilibrium variance [see \cref{eq_var_CCF_film_pbc}].
Owing to the exponential factor in $S(p,k,t)$, the sums and the integral in \cref{eq_Pf_dyn_correl_pbc} are rapidly converging for large momenta $p$ and $k$ and $t>0$. 

We now proceed by analyzing \cref{eq_Pf_dyn_correl_pbc} for $d>2$ in various asymptotic limits, in which exact analytical expressions can be obtained.
In general, the asymptotic behavior is controlled by the exponentials in \cref{eq_Pf_dyn_correl_pbc} [see \cref{eq_S_correl_exp}], such that the integrand contributes significantly only if $q^4 t + q^2\tauLG t\lesssim 1$, where $q^2\equiv p^2 + k_m^2$. This requires $q\lesssim t^{-1/4}$ and $q\lesssim (\tauLG t)^{-1/2}$.
Analogously to \cref{eq_Cdyn_blk_asympt_split}, one can identify two characteristic temporal regimes, depending on which of the two terms dominates. We analyze them separately:

\underline{Case $t\ll \tauLG^{-2}$,} which includes $\tauLG=0$. 
Upon inserting \cref{eq_Cdyn_film_shortT_scal} into \cref{eq_Pf_correl_pbc}, one readily infers that in this regime contributions $\propto \tauLG$ or $\tauLG^2$ to the pressure correlator are subdominant.
A characteristic \emph{long-time} behavior emerges for $t\gg L^4$, provided that also $\tau^{-2}\gg L^4$ holds, i.e., $L^4 \ll t\ll \tauLG^{-2}$.
In this regime, the dominant contribution to the integrand in \cref{eq_Pf_dyn_correl_pbc} stems from small wave numbers, for which $p\ll t^{-1/4}$, $k_j\ll t^{-1/4}$. 
Thus, setting $k_m=k_n=0$, in \cref{eq_Pf_dyn_correl_pbc} all terms involving $\tauLG$ vanish and the integral can be readily evaluated, providing the long-time asymptotic behavior (valid for $d> 2$)
\beq \bra\Delta\Pcal_f\pbc(L^4 \ll t\ll \tauLG^{-2}) \Delta\Pcal_f\pbc(0)\ket \simeq \frac{1 }{2^{(7d-3)/4} \pi^{(d-1)/2} \Gamma(d/4+1/4) A L^2} t^{-(d-1)/4}.
\label{eq_Pf_dyn_correl_lateT_pbc}\eeq
The difference between \cref{eq_Pf_dyn_correl_lateT_pbc} and the corresponding scaling behavior in the bulk [\cref{eq_Pb_dyn_shortT}] stems from the continuum spectrum of modes with small $k$ associated with the infinite transverse ($z$) direction of the bulk system. 
We finally remark that, in a completely finite volume one expects a long-time relaxation behavior different from \cref{eq_Pf_dyn_correl_lateT_pbc} due to the isolated zero mode, which is absent for conserved dynamics \cite{gross_dynamics_2019}.

At \emph{short times} ($t\ll L^4$, still keeping $t\ll \tauLG^{-2}$), instead, modes with $k_j \lesssim t^{-1/4}$ contribute significantly to the integral in \cref{eq_Pf_dyn_correl_pbc}. 
We first focus on dimensions $d\geq 4$, in which case the integrand in \cref{eq_Pf_dyn_correl_pbc} is typically finite for $p\to 0$.
In order to estimate $\bra \Delta\Pcal_f\pbc(t)\Delta\Pcal_f\pbc(0)\ket$ at short times, we replace the sum over the modes by an integral using $\sum_n f(k_n) = (L/(2\pi)) \int \d k   f(k)$.
According to \cref{eq_Pf_correl_pbc}, this replacement can be equivalently performed in the film correlation function in \cref{eq_Cdyn_film_pbc}, rendering [see also \cref{eq_Cdyn_blk_asympt_split}] $\Ccal\pbc(\rvp,z-z',t\ll L^4)\big|_{t\ll \tauLG^{-2}} \simeq \Ccal_b({\rvp,z-z'},t)\big|_{\tauLG=0}$ in terms of the bulk correlation function given in \cref{eq_Cdyn_blk_crit}.
Thus, at short times and for reduced temperatures $\tauLG$ close to bulk criticality, the film and bulk pressure correlations [see \cref{eq_Pb_dyn_shortT}] essentially coincide, i.e.,
\beq \bra \Delta\Pcal_f\pbc(t\ll \min(L^4,\tauLG^{-2})) \Delta\Pcal_f\pbc(0)\ket\big|_{d\geq 4} \simeq \bra \Delta\Pcal_b(t) \Delta\Pcal_b(0)\ket = \frac{H_d}{A} t^{-(d+1)/4} .
\label{eq_Pf_dyn_correl_earlyT_pbc}\eeq 
Note that the notions of a short- and long-time limit are different in the film and in the bulk (compare \cref{sec_dyn_press_bulk}).

For dimensions $d< 4$, the contribution to the integrand in \cref{eq_Pf_dyn_correl_pbc} pertaining to $k_m=0$, $k_n\neq 0$ is diverging for $p\to 0$ (as discussed above) and thus has to be analyzed separately in order to properly determine the short-time behavior.
The scaling behavior of this contribution is found to be
\beq \begin{split} \mathpzc{p} &=  \frac{t^{-d/4} K_{d-1}}{4\pi A L} \int_{\lambda t^{1/4}}^\infty \d P \int_{-\infty}^\infty \d K \, P^{d-4} \frac{P^4 + 2K^2 P^2 + 2K^4}{K^2 + P^2} e^{-K^4-2K^2P^2 - 2P^4} \\
&=  \frac{t^{-d/4}  K_{d-1}}{4\pi A L} \left[ \Gamma(3/4)\int_{\lambda t^{1/4}}^{\sigma} \d P\, P^{d-4} + \mathpzc{R}(\sigma)\right] , 
\end{split}\label{eq_Pf_earlyT_d}\eeq  
where $\sigma \lesssim 1$ is an arbitrary dimensionless parameter introduced to enable the small-$P$ expansion of the integrand, and $\mathpzc{R}(\sigma)$ is a remainder, which is exponentially suppressed and will be omitted henceforth. Since the first equation is independent of $\sigma$, the sum in the second equation must be independent of $\sigma$, too. The required small momentum cutoff $\lambda$ in \cref{eq_Pf_earlyT_d} is proportional to the lateral system size $\lambda\sim A^{-1/(d-1)}$ because $P$ stems from a lateral momentum [see, e.g., \cref{eq_Pf_correl_sample}].
In $d=3$ dimensions \cref{eq_Pf_earlyT_d} leads to 
\beq \mathpzc{p}\overset{d=3}{\simeq} t^{-3/4} \frac{\Gamma(3/4) K_{d-1} }{8\pi^2\, A L} \ln \left(\frac{\sigma}{\lambda t^{1/4}} \right) .
\label{eq_Pf_earlyT_d3}\eeq 
Once this problematic contribution has been removed from \cref{eq_Pf_dyn_correl_pbc}, the continuum approximation leading to \cref{eq_Pf_dyn_correl_earlyT_pbc} can again be applied, such that, in total, the short-time scaling behavior is given by the sum of \cref{eq_Pf_dyn_correl_earlyT_pbc,eq_Pf_earlyT_d}. 
Specifically in $d=3$, this renders
\beq \bra \Delta\Pcal_f\pbc(t\ll \min(L^4,\tauLG^{-2})) \Delta\Pcal_f\pbc(0)\ket\big|_{d=3} \simeq t^{-1} \frac{H_d}{A} + t^{-3/4} \frac{\Gamma(3/4)}{8\pi^2\, A L} \ln \frac{\sigma}{\lambda t^{1/4}} ,
\label{eq_Pf_earlyT_pbc_d3}\eeq 
where the quantity $H_d$ is defined in \cref{eq_Pb_dyn_shortT}.

\underline{Case $t\gg \tauLG^{-2}$.} 
From $t\gg \tauLG^{-2}$, it follows that $t^{-1/4}\gg (\tauLG t)^{-1/2}$. The form of the exponential terms in \cref{eq_Pf_dyn_correl_pbc} then implies that $(\tauLG t)^{-1/2}$ provides an upper bound for $q$ below which the integrand contributes. Thus, in this \emph{long-time} regime, one has $ q^2 / \tauLG \lesssim 1/(t^{1/2}\tauLG) \lesssim 1$ \footnote{This relation follows by using $\tau t^{1/2}\gtrsim 1$ in the inequality $q^2/\tau\lesssim 1/(\tau^2 t)$}, such that the square bracket in \cref{eq_Pf_dyn_correl_pbc} essentially reduces to $\tauLG^2$.
We apply this approximation analogously to the exponentials in \cref{eq_Pf_dyn_correl_pbc} and distinguish the two cases $\tauLG t \gg L^2$ and $\tauLG t \ll L^2$. In the first case we have $\tauLG^2 t\gg\max(1,\tauLG L^2)$, which allows us to set $k_m=k_n=0$. 
Under these assumptions, \cref{eq_Pf_dyn_correl_pbc} renders the following long-time behavior of the film pressure correlations:
\beq \bra \Delta\Pcal_f\pbc(t\gg L^2/\tauLG) \Delta\Pcal_f\pbc(0)\ket \simeq \frac{1}{2^{(3d-1)/2} \pi^{(d-1)/2} A L^2} (\tauLG t)^{-(d-1)/2}.
\label{eq_Pf_lateT_film}\eeq 
For $\tauLG^{-1}\ll \tauLG t\ll L^2$, on the other hand, we can evaluate \cref{eq_Pf_dyn_correl_pbc} by replacing the sum over $k_j$ by an integral. As before, this is equivalent to inserting \cref{eq_Cdyn_larget} into \cref{eq_Pf_correl_pbc_z0}, keeping only the dominant term $(1/2) (\tauLG\, \Ccal\pbc)^2$. This gives the following \emph{intermediate asymptotic} behavior:
\beq \bra \Delta\Pcal_f\pbc(\tauLG^{-2}\ll t\ll L^2/\tauLG) \Delta\Pcal_f\pbc(0)\ket \simeq \frac{1}{2^{3(d+1)/2} \pi^{(d+1)/2} A} (\tauLG t)^{-(d+1)/2}.
\label{eq_Pf_lateT_bulk}\eeq 
Upon increasing $L$, the long-time behavior in \cref{eq_Pf_lateT_film} is gradually shifted towards later times and is replaced by the behavior in \cref{eq_Pf_lateT_bulk}, which coincides with the long-time behavior of the bulk pressure reported in \cref{eq_Pb_dyn_lateT}.
Analogously to \cref{eq_Pf_dyn_correl_lateT_pbc}, the asymptotic long-time behavior in \cref{eq_Pf_lateT_film} is a specific consequence of the continuous spectrum of zero modes in the transverse direction of a thin film.

\begin{figure}[t]\centering
    \subfigure[]{\includegraphics[width=0.48\linewidth]{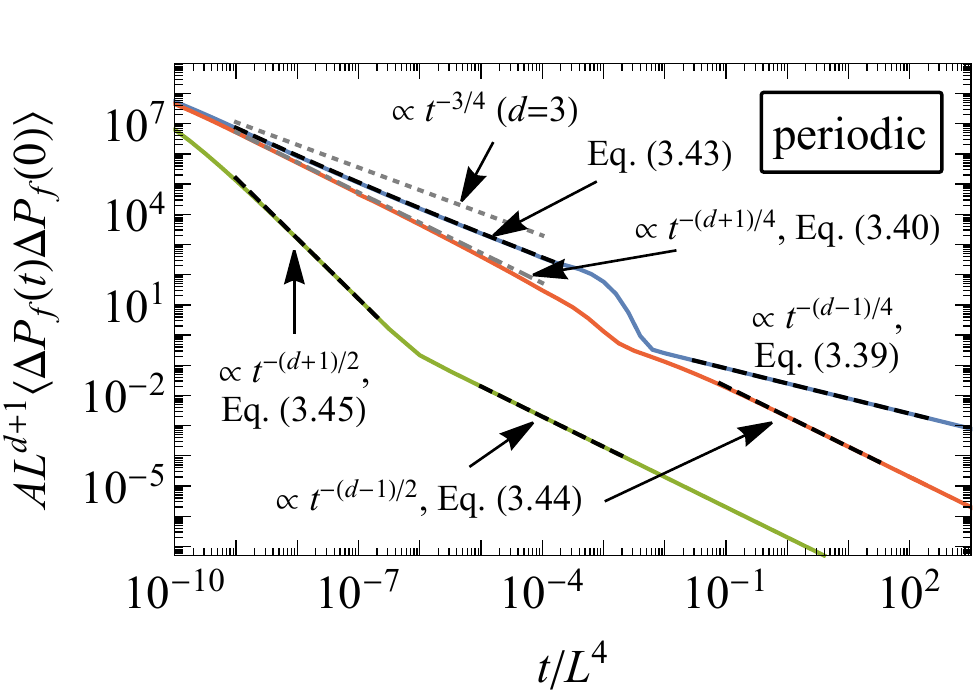}}\qquad
    \subfigure[]{\includegraphics[width=0.48\linewidth]{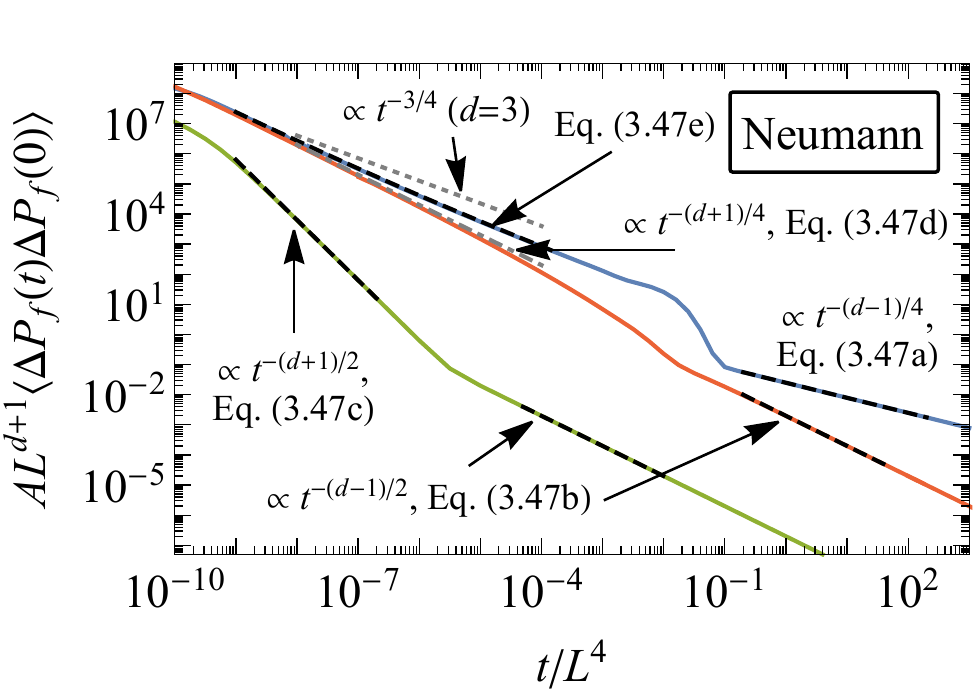}} 
    \caption{Dynamic correlations $\bra \Delta\Pcal_f(t) \Delta\Pcal_f(0)\ket$ of the film pressure for (a) periodic \bcs [\cref{eq_Pf_dyn_correl_pbc}] and (b) Neumann \bcs [\cref{eq_Pf_dyn_correl_Nbc}] for $d=3$ and various temperatures [$L^2 \tauLG=0$ (blue), 10 (red), $10^5$ (green), from top to bottom]. The solid curves are obtained by numerically evaluating \cref{eq_Pf_dyn_correl_pbc,eq_Pf_dyn_correl_Nbc}, while the dashed lines represent the asymptotic laws reported in \cref{eq_Pf_dyn_correl_lateT_pbc,eq_Pf_dyn_correl_earlyT_pbc,eq_Pf_earlyT_pbc_d3,eq_Pf_lateT_film,eq_Pf_lateT_bulk} for panel (a) and in \cref{eq_Pf_dyn_Nbc_asympt} for panel (b). The gray dotted line corresponds to a decay $\propto t^{-3/4}$ (applying to $d=3$), while the gray dash-dotted line represents \cref{eq_Pf_dyn_correl_earlyT_pbc,eq_Pf_dyn_correl_earlyT_Nbc}. These lines are drawn in order to highlight the influence of the logarithmic correction for $d=3$ [see \cref{eq_Pf_earlyT_d3,eq_Pf_earlyT_Nbc_d3}].
    For the evaluation of $\bra \Delta\Pcal_f(t)\Delta\Pcal_f(0)\ket$ at bulk criticality ($\tauLG=0$) we used a low momentum cutoff $\lambda\simeq 10^{-9}/L$ (this specific value is chosen for illustrative purposes). For times shorter than those covered by the plot, the curves corresponding to $L^2\tauLG=10^5$ (green) and $10$ (red) approach the asymptotic critical ($\tau=0$) short-time behavior given in \cref{eq_Pf_earlyT_pbc_d3}. Due to the smaller value of $L^2\tau$, the red curve lies closer to this asymptote and crosses over to the off-critical asymptote [\cref{eq_Pf_lateT_film,eq_Pf_lateT_film_Nbc}] later than the green one. Except for the case of the power law $\propto t^{-3/4}$, the asymptotic laws reported in these plots do not involve a fitting parameter. We note that $\bra \Delta\Pcal_f(t)\Delta\Pcal_f(0)\ket$ has the same dimension as $A^{-1} L^{-(d+1)}$ [see the comment after \cref{eq_Pf_correl}]. }
    \label{fig_PFilmCorrel}
\end{figure}

\subsubsection{Neumann \bcs}
\label{sec_dyncor_Neu}

According to \cref{eq_Pf_correl_Nbc}, the dynamic correlations of the film pressure for Neumann \bcs can be fully expressed in terms of the dynamic OP correlation function $\Ccal\pbc$ [\cref{eq_Cdyn_film_pbc}] for periodic \bcs.
Inserting \cref{eq_Cdyn_film_pbc} into \cref{eq_Pf_correl_Nbc}, we obtain 
\beq\begin{split} 
\bra \Delta\Pcal_f\Nbc(t) &\Delta\Pcal_f\Nbc(0)\ket = \bra\Pcal_f\Nbc(t)\Pcal_f\Nbc(0)\ket - \bra\Pcal_f\Nbc\ket^2 \\
&= \frac{K_{d-1}}{2 A L^2}  \sum_{m,n=-\infty}^\infty  \int_0^\infty \d p\,  S(p,k_m,t) S(p,k_n,t)  p^{d-2}   \left[ 2 (k_n^2 + p^2)(k_n^2+k_m^2) + p^4 + 2\tauLG(2k_n^2+p^2) + \tauLG^2 \right] ,
\end{split}\label{eq_Pf_dyn_correl_Nbc}\eeq 
where the constant $K_d$ and the function $S$ are defined as in \cref{eq_Pf_dyn_correl_pbc}. We emphasize that, although the sum runs over all integers, the wavenumbers $k_{m,n}$ are the ones for Neumann \bcs [\cref{eq_eigenspec_Nbc}]. This is a consequence of the prescription in \cref{eq_Pf_correl_Nbc}, according to which the wavenumbers $k\pbc_{n,m}$ [\cref{eq_eigenspec_pbc}] entering the expressions of $\Ccal\pbc$ are to be evaluated with $2L$ instead of $L$.

As it was the case for periodic \bcs, we find that $\bra(\Delta \Pcal_f\Nbc)^2(t)\ket$ has a weak logarithmic divergence for $d=3$, but is finite for $d>3$ and $t>0$.
Proceeding analogously to the previous subsection, one obtains the following asymptotic time behaviors:
\begin{subequations}
\begin{align}
\bra \Delta\Pcal_f\Nbc(L^4 \ll t\ll \tauLG^{-2}) \Delta\Pcal_f\Nbc(0)\ket &\simeq \frac{1 }{2^{(7d-3)/4} \pi^{(d-1)/2} \Gamma(d/4+1/4) A L^2} t^{-(d-1)/4}, \label{eq_Pf_dyn_correl_lateT_Nbc} \\
\bra \Delta\Pcal_f\Nbc(t \gg \max(\tauLG^{-2},L^2/\tau)) \Delta\Pcal_f\Nbc(0)\ket &\simeq \frac{1}{2^{(3d-1)/2} \pi^{(d-1)/2} A L^2} (\tauLG t)^{-(d-1)/2}, \label{eq_Pf_lateT_film_Nbc} \\
\bra \Delta\Pcal_f\Nbc(\tauLG^{-2}\ll t\ll L^2/\tauLG) \Delta\Pcal_f\Nbc(0)\ket &\simeq \frac{1}{2^{(3d-1)/2} \pi^{(d+1)/2} A} (\tauLG t)^{-(d+1)/2}, \label{eq_Pf_intermT_Nbc}\\
\bra \Delta\Pcal_f\Nbc(t\ll \min(L^4,\tauLG^{-2})) \Delta\Pcal_f\Nbc(0)\ket\big|_{d\geq 4} &\simeq  t^{-(1+d)/4} \frac{H_d\Nbc}{A}, \qquad H_d\Nbc\equiv \Omega_{d-1} \int_0^\infty \d\psi\, F_d\Nbc(\psi) , \label{eq_Pf_dyn_correl_earlyT_Nbc} \\
\bra \Delta\Pcal_f\Nbc(t\ll \min(L^4,\tauLG^{-2})) \Delta\Pcal_f\Nbc(0)\ket\big|_{d=3} &\simeq t^{-1} \frac{H_d\Nbc}{A} + t^{-3/4} \frac{\Gamma(3/4)}{4\pi^2\, A L} \ln \frac{\sigma}{\lambda t^{1/4}} ,  \label{eq_Pf_earlyT_Nbc_d3}
\end{align}\label{eq_Pf_dyn_Nbc_asympt}
\end{subequations}
\hspace{-0.15cm}where the function $F_d\Nbc(\psi)$ follows from \cref{eq_Pf_correl_Nbc} by replacing the film correlator $\Ccal\pbc$ by the bulk correlator $\Ccal_b$ [\cref{eq_Cdyn_blk_crit}]. 
$F_d\Nbc$ is given by a rather lengthy expression and is not stated here.
The meaning of the parameters $\lambda$ and $\sigma$ is the same as in \cref{eq_Pf_earlyT_pbc_d3}.
The quantities $H_d\Nbc$ are finite for $d>2$ and their numerical values are $H_3\Nbc = 2.005 \times 10^{-2}$ and $H_4\Nbc = 2.250 \times 10^{-3}$.
We note that $H_d\Nbc\approx 4 H_d\pbc$ holds only approximately because the expression in \cref{eq_Pf_correl_Nbc} is not a multiple of the one in \cref{eq_Pf_correl_pbc_z0}. Remarkably, this difference, i.e., $H_n\Nbc\neq 4 H_d\pbc$, does not affect most of the other asymptotic scaling laws in \cref{eq_Pf_dyn_Nbc_asympt}; in fact, the behaviors at long times [\cref{eq_Pf_lateT_film_Nbc,eq_Pf_dyn_correl_lateT_Nbc}] coincide with the ones [\cref{eq_Pf_dyn_correl_lateT_pbc,eq_Pf_lateT_film}] pertaining to periodic \bcs. 
An exception is the intermediate asymptotic behavior in \cref{eq_Pf_intermT_Nbc}, which equals four times the one in \cref{eq_Pf_lateT_bulk}.

\subsubsection{Discussion}

By comparing \cref{eq_Pb_dyn_lateT,eq_Pf_dyn_correl_lateT_pbc,eq_Pf_dyn_correl_lateT_Nbc} we infer that, at sufficiently long times, only the contribution of the film pressure is relevant for the correlation function of the CCF [\cref{eq_CCF_correl_dyn}]. Conversely, at short times, the correlation functions of the film and the bulk pressure are both of similar magnitude [see \cref{eq_Pb_dyn_shortT,eq_Pf_dyn_correl_earlyT_pbc,eq_Pf_dyn_correl_earlyT_Nbc}.

\Cref{fig_PFilmCorrel} summarizes the behavior of the film pressure $\bra \Delta\Pcal_f(t) \Delta\Pcal_f(0)\ket$ for (a) periodic and (b) Neumann \bcs as a function of time. The solid curves in \cref{fig_PFilmCorrel} are based on a numerical evaluation of \cref{eq_Pf_dyn_correl_pbc,eq_Pf_dyn_correl_Nbc}, respectively, while the broken lines indicate the various asymptotic power laws obtained in the preceding subsections. Note that these asymptotic predictions do not involve a fitting parameter. In order to regularize the logarithmic infrared divergence of $\bra \Delta\Pcal_f(t) \Delta\Pcal_f(0)\ket$ in $d=3$ at bulk criticality ($\tauLG=0$) [see \cref{eq_Pf_earlyT_pbc_d3,eq_Pf_earlyT_Nbc_d3}], we use a value $\lambda=10^{-9}/L$ for the infrared cutoff. The value of the prefactor is sufficiently small in order to satisfy the condition of considering a thin film, but otherwise arbitrary and chosen for illustrative purposes. For the parameter $\sigma$, which is rather technical and the origin of which is explained in the context of \cref{eq_Pf_earlyT_d}, we use a value $\sigma\simeq 0.1/L$ motivated by numerical considerations. We remark that, due to the logarithmic dependence, using values of these parameters which differ even by an order of magnitude do not noticeably affect the quality of the asymptotic approximations.

\section{Fluctuations of the boundary}
\label{sec_fluct_boundary}

In the preceding section, we have studied static and dynamic pressure correlations at a spatially fixed boundary of the film. We now relax this assumption and consider a responsive, i.e., movable film boundary.

\subsection{Dynamical model}
\label{sec_dyn_model}

We start with a description in terms of the physical time $\tphys$ [see \cref{eq_modelB_bare}] and denote the position of the film boundary by $R(\tphys)$, which initially is at $R_0\equiv R(\tphys=0)=L$ [see \cref{fig_sketch}].
We consider the motion of the boundary to be overdamped and subject to the instantaneous force $\Kcal_z$ [\cref{eq_genforce}] as well as to Gaussian white noise $\eta$ with covariance  $\bra\eta(\tphys)\eta(\tphys\,')\ket = 2\delta(\tphys-\tphys\,')$.
Accordingly, we propose the following Langevin equation for the time evolution of $R$:
\beq \pd_\tphys R(\tphys) = \GAMMA T A \Kcal(R(\tphys),\tphys) + \sqrt{\GAMMA T} \eta(\tphys) = \GAMMA T A\bra\Kcal(R(\tphys))\ket_\phi + \GAMMA T A\Delta\Kcal(R(\tphys),\tphys) + \sqrt{\GAMMA T} \eta(\tphys),
\label{eq_langevin_bndry_bare}\eeq 
where $\GAMMA$ is a mobility coefficient (which has the same dimension as $L^{2}/(T\tphys)$) \footnote{The mobility $\Gamma$ is expected to be proportional to the mobility $\gamma$ of the OP field [\cref{eq_modelB_bare}] \cite{demery_drag_2010, demery_thermal_2011}.}.
The prefactor $T A$ in front of $\Kcal$ arises because we have defined $\Kcal$ as a force per area $A$ of the film at temperature $T$ [see \cref{eq_CCF_Pfilm}].
A reflecting boundary, representing a wall impenetrable to the moving boundary, is taken to be positioned at $R=0$.
In the second equation in \cref{eq_langevin_bndry_bare}, the CCF $\Kcal$ is split according to \cref{eq_CCF_fluct_def} into a mean, $\bra\Kcal\ket_\phi$, and a fluctuating part, $\Delta\Kcal$, (with respect to the OP field). Since we consider thermal equilibrium for the OP field, the mean force $\bra\Kcal\ket$ does not explicitly depend on time.
The white noise $\eta$ accounts for the molecular momentum exchange between the solvent and the surface, which would be present even without a coupling to the OP. 
Upon introducing the rescaled time $t=\gamma \tphys$ as defined in \cref{eq_time_resc}, \cref{eq_langevin_bndry_bare} turns into
\beq \pd_t R(t) =  D  A \bra\Kcal(R(t))\ket_\phi + D  A \Delta\Kcal(R(t),t) + \sqrt{D } \eta(t),\qquad D\equiv T\GAMMA/\gamma,
\label{eq_langevin_bndry_full}\eeq 
where the parameter $D$ is the bare diffusivity with dimension $[D] = [L^{-2}]$. 

The fact, that the stochastic process $\Delta\Kcal(R,\tphys))$ has a nonlinear dependence on $R$, raises the issue of the proper stochastic calculus (e.g., Ito or Stratonovich) to be used in \cref{eq_langevin_bndry_bare} \cite{gardiner_stochastic_2009,mannella_ito_2012}. In principle, this can be addressed by deriving \cref{eq_langevin_bndry_bare} from a more basic model, e.g., by adiabatically eliminating the OP degrees of freedom from a coupled system of Markovian Langevin equations for the boundary and the fluid medium \cite{gardiner_stochastic_2009,pavliotis_stochastic_2014,sancho_brownian_2011,dean_diffusion_2011,demery_perturbative_2011,demery_diffusion_2013,volpe_effective_2016}. 
While this is beyond the scope of the present study, it turns out that, in the Markovian limit, the variance of $\Delta\Kcal$ is approximately independent of $R$, rendering the choice of the stochastic calculus to be immaterial for the present analysis (see \cref{sec_Markov_approx}).

\Cref{eq_langevin_bndry_full} describes the random motion of $R$ in the Casimir potential associated with $\bra\Kcal\ket_\phi$, subject to the Brownian noise $\eta$ [see \cref{eq_langevin_bndry} below] and an additional non-Markovian noise $\Delta\Kcal$. 
While \cref{eq_langevin_bndry_full} resembles previously proposed Langevin equations for inclusions in critical media \cite{demery_perturbative_2011,dean_diffusion_2011,demery_diffusion_2013,gross_dynamics_2021}, here we assume that the presence of a movable boundary does not modify the equation of motion of the OP [\cref{eq_modelB_resc}]. In fact, the only effect of the boundary on the OP is to impose \bcs [see \cref{eq_BCs}]. 
Accordingly, the noise provided by $\Delta\Kcal$ is not balanced by a corresponding friction term, implying that detailed balance is not satisfied by the model defined by \cref{eq_langevin_bndry_full,eq_modelB_resc}.
This is also reflected by the fact that the distribution of the boundary position does not approach the expected equilibrium form at long times (see \cref{app_equil_PDF}). 
We thus refer to our model as having a ``passive boundary''.
For a more adequate description of the dynamics of a colloidal particle in a critical fluid in thermal equilibrium \cite{hertlein_direct_2008,gambassi_critical_2009,maciolek_collective_2018,magazzu_controlling_2019}, detailed balance has to be re-established \footnote{In addition to the limitations of the model spelled out here, experimentally, the surface of the colloid and the wall typically exhibit strong adsorption of the OP. A refined theory would thus not only have to account for the non-planar surface of the colloid, but also consider the so-called $(+\pm)$ \bcs \cite{gambassi_critical_2009} instead of periodic or Neumann ones.}. This can be accounted for by either adding appropriate coupling terms to the OP dynamics in \cref{eq_modelB_resc} (see Ref.\ \cite{gross_dynamics_2021}), or by suitably modifying \cref{eq_langevin_bndry_full} in order to satisfy the fluctuation-dissipation theorem. We will return to the latter approach in \cref{sec_Markov_approx} below. 
However, compared with the models in Refs.\ \cite{demery_perturbative_2011,dean_diffusion_2011,demery_diffusion_2013,gross_dynamics_2021}, a distinctive feature of the present model is that it allows one to enforce strict no-flux \bcs at the location of a boundary.

The mean equilibrium CCF (per area $A$ and temperature $T$) has the form
\beq \bra\Kcal(R)\ket_\phi = R^{-d} \Xi(R/\xi),
\label{eq_CCF_mean}\eeq 
where $\Xi$ is a universal scaling function depending on the boundary conditions. Explicit expressions within the Gaussian approximation are reported in, e.g., Refs.\ \cite{krech_free_1992, gross_statistical_2017}. At criticality (i.e., $\xi\to\infty$), one has $\Xi(0)<0$ for periodic and Neumann \bcs [see \cref{eq_mean_CCF}].
At short distances, the electrostatic repulsion between the fixed wall and the mobile boundary must be taken into account \cite{gambassi_critical_2009,israelachvili_intermolecular_2011}, thereby regularizing also the singularity of the CCF [\cref{eq_CCF_mean}] appearing within the continuum description in the limit $R\to 0$.
Phenomenologically, this electrostatic repulsion can be described by the potential (defined, analogously to the CCF, per area $A$ and temperature $T$) \cite{hertlein_direct_2008}
\beq V\st{es}(z) = \alpha^{-(d-1)} e^{-z/\ell_D},
\label{eq_V_elstat}
\eeq 
where $\alpha$ is an effective parameter (of the same dimension as $L$) and $\ell_D$ denotes the Debye screening length.
Introducing formally the Casimir potential $V\st{C}(z)$ associated with the mean CCF via $\bra\Kcal(z)\ket_\phi = -\d V\st{C}(z) / \d z$, we define an effective potential 
\beq \Ucal(z) = V\st{es}(z) + V\st{C}(z),
\label{eq_eff_pot}\eeq 
which gives rise to a force $-\Ucal'(z)$ replacing $\bra\Kcal(z)\ket_\phi$ in \cref{eq_langevin_bndry_full}. 
Accordingly, \cref{eq_langevin_bndry_full} is replaced by the Langevin equation
\beq \pd_t R(t) = -D A\, \Ucal'(R(t)) +  D A \Delta\Kcal(R(t),t) + \sqrt{D} \eta(t).
\label{eq_langevin_bndry}\eeq 

The effective potential $\Ucal(z)$ has a minimum at a finite distance $z_0$ from the wall. 
In order to facilitate an analytical treatment, we consider its quadratic approximation around its minimum: 
\beq \Ucal(z) \simeq \onehalf \kappa (z-z_0)^2.
\label{eq_eff_pot_harm}\eeq 
An explicit expression for the (temperature dependent) parameter $\kappa$ (which has the same dimension as $L^{-d-1}$) can in principle be obtained from \cref{eq_CCF_mean,eq_V_elstat}, but is not necessary in the following.
As a further simplification, we focus on the limit in which the dynamics of the field $\phi$ is faster than that of the position $R$ (adiabatic approximation). 
Since within the Gaussian model B \cite{hohenberg_theory_1977} the relaxation time of a critical OP fluctuation grows $\propto \Lcal^4$, where $\Lcal$ is the characteristic system size, the adiabatic approximation requires that both the film and the ``bulk'' part of the system are finite (see \cref{fig_sketch}).  
Still, we take the bulk to be sufficiently large so that a continuum approximation is appropriate. 
Requiring $\Lcal$ to be a measure of the largest length scale in the system (which implies also $\Lcal\gtrsim A^{1/(d-1)}$) allows us to introduce a dimensionless ``adiabaticity'' parameter
\beq \chi \equiv D \Lcal^2.
\label{eq_adiab_param}\eeq
In terms of $\chi$, \cref{eq_langevin_bndry_full} can be expressed as  
\beq \pd_t R(t) =  -\chi \frac{A \Ucal'(R(t))}{\Lcal^2} + \chi \frac{A \Delta\Kcal(R(t),t)}{\Lcal^2} + \sqrt{\chi} \frac{\eta(t)}{\Lcal}.
\label{eq_langevin_harm_chi}\eeq 
Thus the adiabatic limit requires $\chi\ll 1$ and facilitates a perturbative solution of \cref{eq_langevin_harm_chi}.
For typical experimental conditions of a colloid immersed in a critical solvent, the condition $\chi\ll 1$ is fulfilled \cite{magazzu_controlling_2019}.

\subsection{Boundary located in the bulk}

Here, we assume here that $R(t)$ is sufficiently far from the wall such that the effective force  $\Ucal'$ in \cref{eq_langevin_bndry} can be neglected.
Accordingly, we have a bulk-like system on both sides of the boundary.
Integrating \cref{eq_langevin_bndry} in time renders
\beq R(t) - R(0) = D A \int_0^t \d s \Delta\Kcal(R(s),s)  + B(t) ,
\label{eq_langevin_int_blk}\eeq 
where $B(t)\equiv \sqrt{D}\int_0^t\d s\, \eta(s)$ is a Brownian motion with
\beq \bra B(t)\ket =0\qquad \text{and}\qquad \bra B(t) B(t') \ket = 2 D \min(t,t').
\label{eq_BM_correl}\eeq 
Upon squaring both sides of \cref{eq_langevin_int_blk} and averaging over the distribution of the position $R$ and the equilibrium configurations of the field $\phi$ [denoted by $\bra\ldots\ket = \bra \bra \ldots\ket_\phi\ket_R$, where $\bra\ldots\ket_\phi \equiv \int\Dcal\phi \ldots P\st{eq}$ with $P\st{eq}$ reported in \cref{eq_eq_dist}], one finds the mean-squared displacement (MSD) 
\beq \bra \left[R(t) - R(0)\right]^2\ket = 2 D t + 2 D^2 A \Mcal_b(t)
\label{eq_msd_R_blk}\eeq 
with 
\beq \Mcal_b(t) \equiv \frac{A}{2} \int_0^t\d s \int_0^t \d s'  \bra \Delta\Kcal(R(s),s) \Delta\Kcal(R(s'),s')\ket ,
\label{eq_msd_CCF_fluct_blk}\eeq 
while the cross-term $\int_0^t \d s  \bra \bra \Delta\Kcal(R(s),s)\ket_\phi B(t)\ket_R $ vanishes because $\bra\Delta\Kcal\ket=0$ [see \cref{eq_CCF_fluct_def}]. 
The quantity $\Mcal_b$ represents the fluctuation contribution to the MSD, which, according to \cref{eq_fluct_CCF_correl}, is determined in the present case by the bulk pressure correlations:
\beq \Mcal_{b}(t) = A \int_0^t\d s \int_0^t \d s'\,  \bra \Delta \Pcal_{b}(R(s),s) \Delta\Pcal_{b}(R(s'),s')\ket .
\label{eq_msd_Pb}\eeq 
The factor 2 in \cref{eq_msd_R_blk} arises because the bulk pressure acts on both sides of the boundary, while in \cref{eq_msd_CCF_fluct_blk} the prefactor $A$ is introduced in order to render $\Mcal$ independent of $A$.
We note that, due to the presence of $\Delta\Kcal$, the process described by \cref{eq_langevin_bndry_full} is highly nonlinear. 

In order to evaluate \cref{eq_msd_CCF_fluct_blk}, we note that in \cref{eq_langevin_harm_chi} the OP-induced noise $\Delta\Kcal$ is of higher order in $\chi$ than the white noise $\eta$. Accordingly, solving \cref{eq_langevin_harm_chi} (assuming, as before, $\Ucal'=0$) in the regime $\chi\ll 1$ to first order in $\chi$ yields a free Brownian motion for $R(t)$: 
\beq R(t)\simeq R(0) + B(t),
\label{eq_R_BM_approx0}\eeq 
which is to be used in \cref{eq_msd_Pb} [see \cref{sec_shorttime_MSD} below].
The required two-time joint probability distribution for a free Brownian process is given by \cite{gardiner_stochastic_2009}
\begin{multline} p(R_s, R_{s'}, R_0, s, s') = \frac{1}{4\pi D \sqrt{s s' - \min(s,s')}}  \exp\Big\{ -\frac{1}{4 D [s s' - \min(s,s')]} \\ \times \big[  s' (R_s - R_0)^2 +  s (R_{s'} -R_0)^2 - 2\min(s,s') (R_s-R_0)(R_{s'}-R_0) \big] \Big\},
\label{eq_jointPDF_BM}\end{multline} 
which renders the following variants of the characteristic function: 
\begin{subequations}
\begin{align}
\big\bra e^{ \im Q(R(s) -  R(s'))} \big\ket &= \int_{-\infty}^\infty \d R_1 \int_{-\infty}^\infty \d R_2\, p(R_1,R_2,R_0,s,s') e^{\im Q (R_1-R_2)}  = e^{-D Q^2 |s-s'|}, \label{eq_charfnc_BM_correl_minus}\\
\big\bra e^{ \im Q(R(s) +  R(s'))} \big\ket &= e^{2\im Q R_0 -D Q^2 (s+s'+2\min(s,s'))}. \label{eq_charfnc_BM_correl_plus}
\end{align}\label{eq_charfnc_BM_correl}
\end{subequations}

\subsection{Boundary located near a wall}
\label{sec_langevin_nearwall}

Here, we retain the effective force $\Ucal'$ in \cref{eq_langevin_bndry,eq_langevin_harm_chi}.
Within the regime $\chi\ll 1$, the fluctuating contribution of the CCF $\Delta\Kcal$ is subdominant compared to the white noise $\eta$, while the deterministic term related to $\Ucal'$ has to be regarded as formally to be of the same order in $\chi$ as $\eta$. (This assignment is also applied, e.g., upon solving a standard Langevin equation of a Brownian particle \cite{gardiner_stochastic_2009}.)
Accordingly, for $\chi\ll 1$, the leading order solution $R(t)$ of \cref{eq_langevin_harm_chi} [based on the assumption formulated in \cref{eq_eff_pot_harm}] is an Ornstein-Uhlenbeck process \cite{gardiner_stochastic_2009}.
However, the calculation of an average as in \cref{eq_msd_CCF_fluct_blk} over this process leads to intractable expressions.
We therefore solve \cref{eq_langevin_harm_chi} instead by applying two complementary approaches: in one (see \cref{sec_directint}), we formally integrate \cref{eq_langevin_harm_chi} in time and focus on short times, in which case the Ornstein-Uhlenbeck process reduces to a free Brownian motion.
In a second approach (see \cref{sec_Markov_approx}), we determine a Markovian approximation of $\Delta\Kcal$ in the adiabatic limit $\chi\ll 1$. Although the second approach holds at all times, we use it specifically to study the long-time behavior.

\subsubsection{Direct integration of \cref{eq_langevin_harm_chi}}
\label{sec_directint}

Upon integrating \cref{eq_langevin_harm_chi} [or, equivalently, \cref{eq_langevin_bndry}] in time one obtains the formal solution
\beq R(t) - R(0) = D A \int_0^t \d s \left[- \Ucal'(R(s)) + \Delta\Kcal(R(s),s)\right] + B(t) ,
\label{eq_langevin_int}\eeq 
with the Brownian motion $B(t)$ [\cref{eq_BM_correl}].
Upon squaring both sides of \cref{eq_langevin_int} and averaging over the (yet unknown) distribution of the boundary position and of the OP field, we obtain 
\begin{multline} \bra \left[R(t) - R(0)\right]^2\ket = 2 D t - D A \int_0^t\d s\, \bra \Ucal'(R(s)) B(t)\ket_R  \\ + (D A)^2 \int_0^t\d s \int_0^t \d s' \big[ \bra\Ucal'(R(s)) \Ucal'(R(s'))\ket + \bra \Delta\Kcal(R(s),s) \Delta\Kcal(R(s'),s')\ket \big] ,
\label{eq_msd_R}\end{multline} 
while, within the approximation eventually considered in \cref{eq_R_BM_approx} below, the cross-term \\ $\int_0^t \d s \left[ (DA)^2 \bra \Ucal'(R(s)) \bra\Delta\Kcal(R(s),s)\ket_\phi\ket_R  + D A \bra \bra \Delta\Kcal(R(s),s)\ket_\phi B(t)\ket_R \right]$ vanishes since $\bra\Delta\Kcal\ket=0$ [see \cref{eq_CCF_fluct_def}] and $R(s)$ is unaffected by $\phi$.
The fluctuating contribution to the MSD is represented by the last term in \cref{eq_msd_R}, which, according to \cref{eq_fluct_CCF_correl}, consists of the correlations of the film and the bulk pressure:
\beq A \int_0^t\d s \int_0^t\d s' \bra \Delta\Kcal(R(s),s) \Delta\Kcal(R(s'),s')\ket = \Mcal_{f}(t) + \Mcal_{b}(t),
\label{eq_msd_CCF_fluct}\eeq 
where $\Mcal_b$ is given by \cref{eq_msd_Pb} and
\beq \Mcal_{f}(t) \equiv A \int_0^t\d s \int_0^t \d s'\,  \bra \Delta \Pcal_{f}(R(s),s) \Delta\Pcal_{f}(R(s'),s')\ket .
\label{eq_msd_Pf}\eeq 

In the limit $\chi\ll 1$, the leading order solution of \cref{eq_langevin_harm_chi} is an Ornstein-Uhlenbeck process, which, at short times $t\ll 1/(DA\kappa)$, reduces to a free Brownian motion [see \cref{eq_BM_correl}]: 
\beq R(t)\simeq R(0) + B(t),\qquad R(0) = z_0 \qquad (t\ll 1/(DA\kappa)).
\label{eq_R_BM_approx}\eeq 
Note that, in order to to simplify the calculation, we consider the film boundary to be initially located at the potential minimum $z_0$.
Upon using \cref{eq_R_BM_approx,eq_jointPDF_BM,eq_charfnc_BM_correl}, the integrals involving $\Ucal'(z) = \kappa (z-z_0)$ in \cref{eq_msd_R} can be readily evaluated, yielding in the short-time limit
\beq\begin{split} \bra \left[R(t) - R(0)\right]^2\ket 
&\overset{t\ll 1/(DA\kappa)}{\simeq} 2 D t  + D^2 A \left[ \Mcal_f(t) + \Mcal_b(t)\right] + \Ocal(\chi^3). 
\label{eq_msd_R_eval}\end{split}\eeq

\subsubsection{Markovian approximation of $\Delta\Kcal$}
\label{sec_Markov_approx}

Here, we seek an alternative solution of \cref{eq_langevin_harm_chi} which is valid for all times.
In order to identify a suitable approximation, we define a new time variable $\hat t$ via 
\beq t=\hat t / \chi
\label{eq_time_chi}\eeq 
and introduce the quantities $\hat R(\hat t\,) = R(\hat t/\chi)$, $\hat\eta(\hat t\,) = \frac{1}{\sqrt{\chi}} \eta(\hat t/\chi)$ [with $\bra \hat\eta(\hat t\,)\hat \eta(\hat t\,') = 2 \delta(\hat t-\hat t\,')$]. In terms of them, \cref{eq_langevin_harm_chi} becomes
\beq \pd_{\hat t} \hat R(\hat t\,) = \frac{A \bra\Kcal(\hat R(\hat t\,))\ket_\phi}{\Lcal^2} + \frac{A \Delta\Kcal(\hat R(\hat t\,),\hat t/\chi)}{\Lcal^2} +  \frac{\hat \eta(\hat t\,)}{\Lcal},
\label{eq_langevin_bndry_resc}\eeq 
where, using the time-translation invariance of the CCF fluctuations, the correlation function of the field-induced noise $\Delta\Kcal$ [see \cref{eq_fluct_CCF_correl,eq_Pf_correl}] is given by
\beq \hat\Ncal(\hat R, \hat R\hspace{0.5pt}', \hat s-\hat s\hspace{0.5pt}') \equiv \bra\Delta\Kcal(\hat R, 0) \Delta\Kcal(\hat R\hspace{0.5pt}', \sfrac{(\hat s - \hat s\hspace{0.5pt}')}{\chi})\ket .
\label{eq_field_noise_var}\eeq

In the limit $\chi\ll 1$, one expects from \cref{eq_langevin_bndry_resc,eq_field_noise_var} that $\Delta\Kcal$ decorrelates rapidly in time, justifying a Markovian approximation for $\hat\Ncal$. 
Furthermore, we assume $\Delta\Kcal$ to be Gaussian, such that cumulants higher than the second one reported in \cref{eq_field_noise_var} are zero \footnote{The noise $\Delta\Kcal$ is, in general, not Gaussian, because it follows by applying the nonlinear transformations, given by Eqs.\ \eqref{eq_CCF_Pfilm}, \eqref{eq_Pfilm}, \eqref{eq_Tzz_dyn}, and \eqref{eq_CCF_fluct_def}, to the (Gaussian) field $\phi$ defined by the stochastic process in \cref{eq_modelB_resc}. Accordingly, the assumption of $\Delta\Kcal$ being Gaussian is necessarily an approximation. We recall that, at least in the static case, Monte Carlo simulations have confirmed the Gaussian character of $\Delta\Kcal$ \cite{dantchev_critical_2004}.}. 
From standard adiabatic elimination techniques \cite{gardiner_stochastic_2009,stratonovich_topics_1963} it follows that $\Delta\Kcal$ can be approximated by a Markovian noise $\Delta\Kcal_M$ with the variance 
\beq \bra \Delta\Kcal_M(R(t),t) \Delta\Kcal_M(R(t'),t')\ket = 2\Ncal(R(t))\delta(t-t'),\qquad \Ncal(R) \equiv \int_{0}^\infty \d u\, \hat\Ncal( R,  R, u).
\label{eq_Knoise_Markov}\eeq
The resulting Stratonovich-type Langevin equation \cite{gardiner_stochastic_2009,stratonovich_topics_1963} is given by
\beq \pd_t R(t) = -D A\, \Ucal'(R(t)) +  D A \Delta\Kcal_M(R(t),t) + \sqrt{D} \eta(t)
\label{eq_langevin_Markov}\eeq 
and corresponds to the following Fokker-Planck equation for the probability density $P(R,t)$ of the position $R$ of the movable boundary:
\beq \pd_t P(R,t) = \pd_R \left[ \left( D A\, \Ucal'(R) - \frac{1}{2} (D A)^2\Ncal'(R) \right) P(R,t)\right]  + \pd^2_R\left[ \left(D +  (D A)^2\Ncal(R) \right) P(R,t)\right],
\label{eq_FPE_Markov_full}\eeq 
where the term involving $\Ncal'(R)$ is a consequence of the Stratonovich stochastic calculus, also referred to as ``spurious'' drift \cite{gardiner_stochastic_2009}.
The expression for the noise amplitude $\Ncal(R)$ (which has the same dimension as $L^{4-2d}$) is obtained by using \cref{eq_Cdyn_film_pbc,eq_C_film_Neu,eq_fluct_CCF_correl,eq_Pf_correl} in \cref{eq_Knoise_Markov}:
\beq \Ncal(R) = \frac{K_{d-1}}{A } \int_0^\infty \d p\, \left[ \frac{1}{L^2} \sum_{m,n=-\infty}^\infty\, \Qcal\ut{(p,N)}\left(k_m\ut{(p,N)},k_n\ut{(p,N)},p\right) + \int_{-\infty}^\infty \frac{\d k_m}{2\pi} \int_{-\infty}^\infty \frac{\d k_n}{2\pi} \Qcal_b(k_m,k_n,p) \right],
\label{eq_Knoise_Markov_var}\eeq 
where the wavenumbers used in the sum are defined in \cref{eq_eigenspec_pbc} [see also the comment after \cref{eq_Pf_dyn_correl_Nbc}]; $K_d$ is defined after \cref{eq_Pf_dyn_correl_pbc}, and  
\begin{subequations}
\begin{align}
\Qcal_b = \Qcal\pbc  &\equiv   p^{d-2}  \frac{3k_m^2 k_n^2 +2k_n^4 + 2(k_m^2+k_n^2)p^2 + p^4}{(k_m^2+p^2)(k_n^2+p^2)(k_m^4 + k_n^4 + 2(k_n^2+k_m^2)p^2+2p^4)}, \label{eq_Q_Markov_pbc}\\
\Qcal\Nbc &\equiv   p^{d-2} \frac{1}{8 \left(k_m^2+p^2\right) \left(k_n^2+p^2\right) \left[2 p^2 \left(k_m^2+k_n^2\right)+k_m^4+k_n^4+2 p^4\right]} \nonumber \\ &\qquad
\Big\{k_m^2 \Big[2 p^2 \left(1+e^{2 \im R k_m}\right) \left(1+e^{2 \im R k_n}\right)+k_n^2 \left(3 e^{2 \im R \left(k_m+k_n\right)}+e^{2 \im R k_m}+e^{2 \im R   k_n}+3\right)\Big] \nonumber \\ &\qquad -2 k_m k_n \left(2 k_n^2+p^2\right) \left(1+e^{2 \im R \left(k_m+k_n\right)}\right)+\left(1+e^{2 \im R k_m}\right) \left(2 p^2 k_n^2+2 k_n^4+p^4\right)   \left(1+e^{2 \im R k_n}\right)\Big\}, \label{eq_Q_Markov_Nbc}
\end{align}\label{eq_Q_Markov}
\end{subequations}
\hspace{-0.11cm}for $\tauLG=0$ \footnote{We recall that $\Qcal\Nbc$ is obtained by using \cref{eq_C_film_Neu}, according to which $\Ccal\pbc$ is evaluated as it would be for a film of thickness $2L$ instead of $L$. As a consequence, the wave numbers in \cref{eq_Q_Markov_Nbc} take the same form as those for Neumann \bcs [\cref{eq_eigenspec_Nbc}], but the sum in \cref{eq_Knoise_Markov_var} runs from $-\infty$ to $\infty$}. 
The expressions of these quantities for $\tauLG \neq 0$ are rather lengthy and thus they are not stated here. In general, the integral over $p$ diverges both at its lower (0) and upper ($\infty$) limits and thus has to be regularized, which will be discussed below.
Symmetry arguments show that $\sum_{m,n} \Qcal\Nbc$ is real valued.  
Note that $\Qcal\Nbc$ is also a function of $R$, while $\Qcal_b$ and $\Qcal\pbc$ are not.
However, a numerical analysis reveals that $\Qcal\Nbc$ is only weakly depending on $R$ (provided $L\lambda\ll 1$), which allows us to make the approximation $\Qcal\Nbc(R) \simeq \Qcal\Nbc(R=0)$ so that $\Ncal\Nbc(R)\simeq \Ncal\Nbc(R=0)$. This renders $\Ncal$ to be independent of $R$ for all \bcs considered here. Accordingly, \cref{eq_FPE_Markov_full} can be approximated by
\beq \pd_t P(R,t) = \pd_R \left[ D A\, \Ucal'(R)  P(R,t)\right]  + \pd^2_R\left[ \left(D + \ (D A)^2\Ncal \right) P(R,t)\right] .
\label{eq_FPE_Markov}\eeq 
\Cref{eq_FPE_Markov} describes a passive boundary at position $R$ moving in the harmonic potential $\Ucal$ [\cref{eq_eff_pot_harm}] under the influence of the white noise of variance $2D$ [see \cref{eq_langevin_bndry_full}] and a Markovian noise of variance $2(D A)^2 \Ncal$ [see \cref{eq_Knoise_Markov_var}].

Next, we determine the Fokker-Planck equation (FPE) of a boundary in a system with detailed balance. To this end, we assume that the corresponding FPE takes the same form as in \cref{eq_FPE_Markov}, up to a different (but still spatially constant) mobility $\Gamma\eff$, i.e., 
\beq \pd_t P(R,t) = \pd_R \left[ \Gamma\eff A\, \Ucal'(R)  P(R,t)\right]  + \pd^2_R\left[ \left(D + (D A)^2\Ncal \right) P(R,t)\right] .
\label{eq_FPE_Markov_therm}\eeq 
We now require that the fluctuations of the boundary are described by the equilibrium distribution (see \cref{app_equil_PDF})
\beq P\st{eq}(R) = \frac{1}{\Zcal}e^{-A\Ucal(R)}.
\label{eq_FPE_eqpdf}\eeq 
In order that the steady-state solution of \cref{eq_FPE_Markov_therm} agrees with $P\st{eq}$, the following relation must hold \cite{gardiner_stochastic_2009}:
\beq \Gamma\eff =  D+(DA)^2\Ncal.
\label{eq_mobility}\eeq 
This can be regarded as a consequence of the fluctuation-dissipation theorem \cite{kubo_fluctuation-dissipation_1966}.  
We shall refer to a boundary described by the model in \cref{eq_FPE_Markov_therm,eq_mobility} as ``thermalized''.
Importantly, asymptotically at short times, the behaviors of the MSD resulting from \cref{eq_FPE_Markov,eq_FPE_Markov_therm} are identical and independent of the mobility (see \cref{sec_shorttime_MSD}).
The derivation of the effective mobility of a boundary from a more rigorous approach (following, e.g., Refs.\ \cite{demery_drag_2010,demery_thermal_2011,gross_dynamics_2021}) is left for future studies.

Denoting the second term in the square bracket in \cref{eq_Knoise_Markov_var} by $\mcal_{\,b}$ (where $b$ stands for bulk), power counting (taking into account the integrals over $k_m$ and $k_n$) implies the asymptotic behaviors
\begin{subequations}
\begin{align}
\mcal_b(p\to 0,\infty)\big|_{\tauLG=0} &\propto p^{d-4}, \\
\mcal_b(p\to 0)\big|_{\tauLG>0} &\propto p^{d-2}, \\
\mcal_b(p\to \infty)\big|_{\tauLG>0} &\propto p^{d-4}.
\end{align}\label{eq_Knoise_Markov_mb_lims}
\end{subequations}
\hspace{-0.15cm}The first term in the square brackets in \cref{eq_Knoise_Markov_var}, denoted by $\mcal_f$ (where $f$ stands for film), is dominated for small $p$ by the contribution for $k_m=k_n=0$ [see \cref{eq_eigenspec_pbc,eq_eigenspec_Nbc}] and follows the asymptote
\begin{subequations}
\begin{align}
\mcal_f(p\to 0)\big|_{\tauLG=0} &\propto p^{d-6}, \label{eq_Knoise_Markov_film_int_lowp}\\
\mcal_f(p\to 0)\big|_{\tauLG>0} &\propto \tauLG^{-1} p^{d-4},
\end{align}\label{eq_Knoise_Markov_mf_lims}
\end{subequations}
\hspace{-0.15cm}independently of the \bcs.
For large $p$, a numerical analysis indicates a behavior as in \cref{eq_Knoise_Markov_film_int_lowp}:
\beq \mcal_f(p\to \infty) \propto p^{d-6},
\eeq 
independently of $\tauLG$.
In order to regularize the integral over $p$ in \cref{eq_Knoise_Markov_var}, we replace its lower and upper cutoff by wavenumbers $\lambda$ and $\Lambda$, respectively.
Accordingly, within the Markovian approximation in $d=3$, $\Ncal_{b}\equiv \int_\lambda^\Lambda\d p\, \mcal_b$ diverges logarithmically as $\lambda\to 0$ or $\Lambda\to \infty$. In contrast, $\Ncal_f\equiv \int_\lambda^\Lambda \d p\,\mcal_f$ is finite for $\Lambda\to \infty$, while it diverges $\propto \Lcal^2$ with the characteristic (lateral) system size $\Lcal$ if $\lambda\simeq 1/\Lcal$ is taken for the lower integration limit.

\subsection{Short-time behavior of the MSD}
\label{sec_shorttime_MSD}

Based on the expressions derived above, in the following we discuss the evolution of the mean-squared displacement (MSD) of the movable boundary at short times and elucidate how the fluctuations of the CCF affect it.

\subsubsection{Preliminaries}

The expressions of $\Mcal_f$ [\cref{eq_msd_Pf}] for periodic \bcs and $\Mcal_b$ [\cref{eq_msd_Pb}] for the bulk are structurally similar and can thus be analyzed together.
$\Mcal_f\pbc$ is obtained by inserting the Fourier transform of the correlation function given in \cref{eq_Cdyn_film_pbc} into \cref{eq_Pf_correl_pbc}. 
In contrast to the calculation of the pressure variance at a fixed position discussed in \cref{sec_dyncor_per}, we now have to retain the dependences on the $z$-coordinate (as well as on the second time argument $t'$, which was previously set to 0) and subsequently average over the distribution of the boundary position $z=R$.
Moreover, we cannot use the simplifications stemming from \cref{eq_correl_pd_z0,eq_correl_pd_z0_Nbc}.
In order to keep the expressions tractable, we focus on the bulk critical point ($\tauLG=0$).

We illustrate the calculation of $\Mcal_f$ by considering a typical contribution to the integrand in \cref{eq_msd_Pf} for periodic \bcs [compare \cref{eq_Pf_correl_sample} and see \cref{eq_eigenspec_pbc}]:
\beq\begin{split} &\frac{1}{A} \int \d^{d-1}r\, \big\bra [\pd_\alpha^2 \Ccal\pbc(\rv,R(s)-R(s'),s-s')][\pd_{\alpha'}^2 \Ccal\pbc(\rv,R(s)-R(s'),s-s')] \big\ket \\
&\qquad = \frac{1}{A L^2} \int \d^{d-1}r \sum_{m,n=-\infty}^\infty \int \frac{\d^{d-1} p}{(2\pi)^{d-1}} \int \frac{\d^{d-1} \tilde p}{(2\pi)^{d-1}}   p_\alpha^2 \tilde p_{\alpha'}^2 \big\bra e^{\im(\pv+ \tilde \pv)\cdot \rv + \im [R(s)-R(s')] (k_m + k_n) } \big\ket  S(p,k_m,s-s') S(\tilde p,k_n,s-s') \\
&\qquad = \frac{1}{A L^2} \sum_{m,n=-\infty}^\infty \int \frac{\d^{d-1} p}{(2\pi)^{d-1}}  p_\alpha^2 p_{\alpha'}^2 \big\bra e^{ \im [R(s)-R(s')] (k_m+k_n) } \big\ket S(p,k_m,s-s') S(p,k_n,s-s'),
\end{split}\label{eq_msd_sample_term}\eeq 
where the function $S$ (which depends only on even powers of $p$) is given in \cref{eq_S_correl_exp}.
In order to proceed, we approximate $R(s)$ as a Brownian motion as formulated in \cref{eq_R_BM_approx0,eq_R_BM_approx}. We recall that, in the bulk, this approximation holds at all times, whereas it requires $t\ll 1/(DA\kappa)$ in the case of a film.
Next, we replace in \cref{eq_msd_sample_term}, as well as in all other contributions stemming from \cref{eq_msd_Pf}, the characteristic function by the expression in \cref{eq_charfnc_BM_correl_minus} (setting $Q=k_m+k_n$).
After performing the time integrals, one obtains in total for a film with periodic \bcs
\begin{subequations}
\begin{align}
 \Mcal_f\pbc(t) &= \frac{K_{d-1}}{L^2} \sum_{m,n =-\infty}^\infty \int_0^\infty \d p\, \Gcal\pbc\left(k_n\pbc, k_m\pbc, p, t\right)
\intertext{with }
\Gcal\pbc &= \Bigg[t - \frac{ \left(1-e^{-\Ecal\pbc t}\right)}{\Ecal\pbc}\Bigg] \frac{ p^{d-2} \left[-2 k_m k_n \left(2 k_n^2+p^2\right)+k_m^2 \left(3 k_n^2+2 p^2\right)+2 p^2 k_n^2+2
   k_n^4+p^4\right]}{\Ecal\pbc \left(k_m^2+p^2\right) \left(k_n^2+p^2\right)} ,\label{eq_msd_finalexpr_b} \\
\Ecal\pbc &= 2 D k_m k_n+k_m^2 \left(2 p^2+D\right)+k_m^4+k_n^2 \left(2 p^2+D\right)+k_n^4+2 p^4,
\end{align}
\label{eq_msd_finalexpr}
\end{subequations}
\hspace{-0.1cm}and where $K_d$ is defined in the context of \cref{eq_Pf_dyn_correl_pbc}.
The corresponding expression for $\Mcal_b$ is given by replacing in \cref{eq_msd_finalexpr} the sums over $k_{m,n}$ by integrals according to 
\beq \frac{1}{L} \sum_{j=-\infty}^\infty f(k_j=2\pi j/L) \to \int_{-\infty}^\infty \frac{\d k}{2\pi} f(k)
\label{eq_bulk_repl}\eeq
which holds for any function $f(k)$.

For a film with Neumann \bcs, $\Mcal_f\Nbc$ follows from using \cref{eq_C_film_Neu} in \cref{eq_Pf_correl} and then by proceeding analogously to the calculation which leads to \cref{eq_msd_finalexpr}.
We note that here both variants of the characteristic function in \cref{eq_charfnc_BM_correl} are needed and that the resulting expression has to be evaluated with $2L$ instead of $L$. Since the result is rather lengthy it is not reported here.
With the exception of the short- and long-time limits (see below), $\Mcal_{f,b}$ have to be determined numerically.

\subsubsection{Bulk system}

In order to calculate \cref{eq_msd_finalexpr} for a bulk system, in the integrals we rescale all momenta by $t^{-1/4}$ and regularize the integral over $p$ by replacing its lower and upper integration limits by $\lambda$ and $\Lambda$, respectively.
The resulting expression is independent of $D$ in the limit $t\ll D^{-2}$, which we identify as the short-time regime for the bulk. In this regime one finds the scaling behavior (up to possible logarithmic corrections, see below)
\beq 
\Mcal_{b}(t\ll D^{-2}) \simeq t^{\frac{7-d}{4}} \mathcal{W}
\label{eq_msd_shortT_bulk}\eeq 
with 
\beq \mathcal{W} = \int_{\lambda^*}^{\Lambda^*} \d P\, \mathpzc{w}(P),
\label{eq_msd_shortT_int}\eeq 
where the (dimensionless) integrand behaves as $\mathpzc{w}(P\to 0) \propto P^{d-3}$ and $\mathpzc{w}(P\to\infty) \propto P^{d-4}$. 
The dimensionless integration boundaries $\lambda^*$ and $\Lambda^*$ in \cref{eq_msd_shortT_int} are related to the physical momenta $\lambda$ and $\Lambda$ via $\{\lambda^*,\Lambda^*\} = \{\lambda,\Lambda\} t^{1/4}$.
In $d=3$, the integral in \cref{eq_msd_shortT_int} is finite for $\lambda^* \to 0$ but diverges logarithmically upon increasing the value of the upper cutoff $\Lambda^*\to\infty$.
By numerically determining the integral in $d=3$ with $\lambda=0$, we find  
\beq \Wcal \simeq 0.048 + 0.011 \times \ln (t \Lambda^4).
\label{eq_msd_shortT_blk_d3corr}\eeq 
Accordingly, in $d=3$, the algebraic time dependence in \cref{eq_msd_shortT_bulk} is modified by a weak logarithmic correction.
In $d=2$, instead, a logarithmic correction is induced by the infrared divergence of $\wcal$. (For $d=1$ a separate calculation is required.)

\subsubsection{Film with periodic \bcs}

We now consider a film boundary fluctuating near a fixed wall and note that, for $t\ll L^4$, the exponential in \cref{eq_msd_finalexpr_b} varies mildly over a wide range of momenta. 
Accordingly, one can approximate the sums in $\Mcal_f\pbc$ in \cref{eq_msd_finalexpr} by integrals and, as above, rescale all momenta by $t^{-1/4}$.
As before, we additionally consider the regime $t\ll 1/(DA\kappa)$ and note that, consequently, $t\ll D^{-2}$ is automatically fulfilled owing to the adiabatic approximation $D\ll L^{-2}$.
The short-time regime for $\Mcal_f\pbc$ turns out to coincide with \cref{eq_msd_shortT_bulk}, i.e.,
\beq 
\Mcal_{f}\pbc(t\ll L^4) \simeq t^{\frac{7-d}{4}} \mathcal{W},
\label{eq_msd_shortT_pbc}\eeq 
with $\Wcal$ given by \cref{eq_msd_shortT_int,eq_msd_shortT_blk_d3corr}.

\subsubsection{Film with Neumann \bcs}

A scaling analysis of the expression $\Mcal_f\Nbc$ (not reported here) yields, analogously to \cref{eq_msd_shortT_pbc} and under the assumption $t\ll 1/(DA\kappa)$, the short-time behavior
\beq 
\Mcal_{f}\Nbc(t\ll L^4) \simeq t^{\frac{7-d}{4}} \mathcal{W}\Nbc
\label{eq_msd_shortT_Nbc}\eeq 
with 
\beq \mathcal{W}\Nbc = \int_{\lambda^*}^{\Lambda^*} \d P\, \wcal\Nbc(P) \overset{d=3}{\simeq} 0.087 + 0.02\times \ln (t \Lambda^4).
\label{eq_msd_shortT_int_Nbc}\eeq 
As before, the physical momentum cutoffs $\lambda$ and $\Lambda$ enter via $\{\lambda^*,\Lambda^*\} = \{\lambda, \Lambda\} t^{1/4}$, and the function $\wcal\Nbc(P)$ behaves asymptotically in the same way as $\wcal$ in \cref{eq_msd_shortT_int}.
As it was the case for a bulk system and for the one with periodic \bcs [see \cref{eq_msd_shortT_bulk,eq_msd_shortT_pbc}], the algebraic increase of $\Mcal_f\Nbc(t)$ at small times is modified by a weak, logarithmic time dependence.

\subsubsection{Integrability of the pressure correlations}
\label{sec_integr}

Here, we elucidate the connection between the MSD and the pressure correlations $\bra\Delta\Pcal(t)\Delta\Pcal(0)\ket$ as determined in \cref{sec_dyn_press} (see, in particular, \cref{fig_PFilmCorrel}).
Instead of using \cref{eq_msd_finalexpr}, the MSD [\cref{eq_msd_R_blk,eq_msd_R_eval}] can be obtained in an approximate way by noting that the probability distribution $p(R_s,R_{s'}, R_0,s,s')$ in \cref{eq_jointPDF_BM} is strongly peaked for sufficiently short times $s$, $s'$ and exhibits the maximal statistical weight for $R_s\approx R_{s'}$.
Accordingly,  in the short-time regime, the pressure correlations entering into \cref{eq_msd_CCF_fluct_blk,eq_msd_Pb,eq_msd_Pf} can be estimated as 
$\bra\Delta\Pcal(R(s),s)\Delta\Pcal(R(s'),s')\ket \sim \bra\Delta\Pcal(s-s')\Delta\Pcal(0)\ket$ 
in terms of the pressure correlations at a \emph{fixed} boundary position as calculated in \cref{sec_dyn_press}.
Within this approximation, an algebraic behavior $\bra\Delta\Pcal(t)\Delta\Pcal(0)\ket\propto t^{-\alpha}$ implies $\Mcal\propto t^{2-\alpha}$, provided $\alpha<2$. For $\alpha\geq 2$, instead, the pressure correlations are not integrable.
Specifically, for the short-time expressions reported in \cref{eq_Pb_dyn_shortT,eq_Pf_dyn_correl_earlyT_pbc,eq_Pf_dyn_correl_earlyT_Nbc} one has $\alpha=(d+1)/4$ and thus recovers the algebraic behaviors in \cref{eq_msd_shortT_bulk,eq_msd_shortT_pbc,eq_msd_shortT_Nbc}.

\subsection{Long-time behavior of the MSD}
\label{sec_longtime_MSD}

We now turn to a discussion of the MSD of a movable boundary and of the influence of the CCF on it at long times. 

\subsubsection{Bulk system}

The long-time behavior of a boundary in a bulk system is provided by \cref{eq_msd_R_blk} together with \cref{eq_msd_finalexpr} [after making the appropriate replacement indicated in \cref{eq_bulk_repl}].
According to the analysis leading to \cref{eq_msd_shortT_bulk}, a long-time regime emerges for times $t\gg D^{-2}$.
In this limit, the exponential term in \cref{eq_msd_finalexpr_b} turns out to be negligible, and the remaining expression renders a linear time dependence: 
\beq 
\Mcal_{b}(t\gg D^{-2}) \simeq t\, \int_\lambda^\Lambda \d p\, \mathpzc{v}(p).
\label{eq_msd_lateT_bulk}\eeq 
In order to assess the relevance of the cutoffs  $\lambda$ and $\Lambda$, we note that the integrand behaves as $\mathpzc{v}(p\to\infty)\propto p^{d-4}$, while the behavior at small $p$ depends in general on $D$. For $D\to 0$, one has $\mathpzc{v}\propto p^{d-4}$, while at nonzero $D$, a numerical analysis indicates a behavior close to $\mathpzc{v}\propto p^{d-3}$.
In spatial dimension $d=3$ with nonzero $D$, one may thus set $\lambda=0$, while retaining a logarithmic dependence on the microscopic length scale $\varepsilon\propto 1/\Lambda$ via the dimensionless combination $D \varepsilon^2$.
Since the function $\mathpzc{v}$ depends on the diffusivity $D$, in general the calculation of \cref{eq_msd_lateT_bulk} requires a numerical approach (see \cref{sec_msd_discussion} for further discussions).

\subsubsection{Film with periodic \bcs}

In order to determine the long-time behavior of the MSD of a fluctuating boundary in proximity to another wall, we invoke the Markovian approximation discussed in \cref{sec_Markov_approx}, which leads to an Ornstein-Uhlenbeck process for the boundary position [see \cref{eq_langevin_Markov,eq_FPE_Markov,eq_FPE_Markov_therm}].
\Cref{eq_FPE_Markov,eq_FPE_Markov_therm,eq_mobility} imply at long times a static Gaussian distribution with variance \cite{gardiner_stochastic_2009} 
\begin{subequations}
  \label{eq_var_lateT}
    \begin{empheq}[left={\bra [\Delta R]^2\ket_\infty \equiv \bra [\Delta R(t\to\infty)]^2\ket =\empheqlbrace\,}]{align} \displaystyle
	& \frac{1}{\kappa A} + \frac{DA \Ncal}{2\kappa}, &\text{(passive)} \label{eq_var_lateT_pass} \\ \displaystyle
	& \frac{1}{\kappa A}, &\text{(thermalized)} \label{eq_var_lateT_th}
	\end{empheq}
\end{subequations}
which depends on whether the boundary is passive or thermalized; $\Ncal$ is given by \cref{eq_Q_Markov_pbc,eq_Knoise_Markov_var}. 
The expression in \cref{eq_var_lateT_th} follows equivalently from \cref{eq_FPE_eqpdf}, because, in fact, the FPE in \cref{eq_FPE_Markov_therm} has been constructed as to yield this result. This term is determined by the variance of the thermal white noise $\eta$ and the curvature $\kappa$ of the effective potential [see \cref{eq_eff_pot_harm}], which includes the competing contributions of the electrostatic repulsion and of the CCF (the factor $1/A$ arises because the associated force $-\Ucal'$ is defined per area).
We note that the temperature enters this expression only through $\kappa$ via the temperature dependence of the scaling function of the CCF [see \cref{eq_CCF_mean}].
The second term in \cref{eq_var_lateT_pass} stems from the fluctuations of the fluctuation-induced force. For a passive boundary they are not balanced by a corresponding friction term and thus contribute to the steady-state variance.
In contrast, kinetic coefficients such as $D$ in general do not enter into steady-state equilibrium quantities.
 
Specifically for $d=3$ and at bulk criticality ($\tauLG=0$), \cref{eq_var_lateT_pass} leads to 
\beq \bra [\Delta R]^2\ket_{\infty}\ut{pass} \simeq \frac{1}{\kappa A} +  \frac{ D }{\kappa }\left[ 0.46 - 0.042\times \ln(\lambda/\Lambda) + 0.040\times (L\lambda)^{-2} \right],
\label{eq_var_lateT_d3}\eeq 
where $\lambda$ and $\Lambda$ denote small- and large-momentum cutoffs which stem from the integral over the lateral momentum in \cref{eq_Knoise_Markov_var}.
Upon naturally identifying the largest length scale $\Lcal$ in the system as $\Lcal\simeq A^{1/2}$, such that $\lambda\propto A^{-1/2}$, and upon introducing a microscopic (molecular) length $\varepsilon \sim 1/\Lambda$ (specified below), \cref{eq_var_lateT_d3} can be expressed as
\beq \bra [\Delta R]^2\ket_{\infty}\ut{pass} \simeq \frac{1}{\kappa A} \left\{1 +  DA\left[ 0.46 + 0.042\times \ln(A^{1/2}/\varepsilon) + 0.040\times \frac{A}{L^2} \right]\right\}.
\label{eq_var_lateT_d3_estim}\eeq 
We remark that, off criticality, the term proportional to $A/L^2$ in the square brackets disappears owing to the much weaker divergences of the expressions in \cref{eq_Knoise_Markov_mb_lims,eq_Knoise_Markov_mf_lims} for $\tauLG>0$.

\Cref{eq_var_lateT_th} provides a first approximation for the variance of the position of a colloidal particle close to a wall and immersed in a nearly critical solvent in thermal equilibrium, consistent with previous studies \cite{hertlein_direct_2008,magazzu_controlling_2019,maciolek_collective_2018}. 
Experimentally, the typical magnitude of the fluctuations is of the order of $\bra [\Delta R]^2\ket_\infty^{1/2} \approx 10\,\mathrm{nm}$ \footnotetext[200]{This implies a value $\kappa\approx 10^{28}\,\mathrm{m}^{-4}$ [see \cref{eq_eff_pot_harm}].} \cite{hertlein_direct_2008,Note200}.
In order to evaluate \cref{eq_var_lateT_d3_estim}, which is experimentally less relevant due to the underlying passive boundary approximation, we take $A$ to be the cross sectional area of a typical colloidal particle, $A\simeq (1\,\mu\mathrm{m})^2$, and estimate the molecular length scale of the solvent as $\varepsilon\simeq 1\,\mathrm{nm}$, in which case \cref{eq_var_lateT_d3_estim} reduces to $\bra \Delta R^2\ket_\infty\ut{pass} \simeq \frac{1}{\kappa A}(1+0.75\times DA)$.  
Because \cref{eq_adiab_param} (with $\Lcal\simeq A^{1/2}$) implies $DA\ll 1$ in the adiabatic limit, the contribution to $\bra [\Delta R]^2\ket_\infty\ut{pass}$ stemming from the fluctuations of the CCF is subdominant relative to those stemming from the thermal white noise in \cref{eq_langevin_Markov}.

\subsubsection{Film with Neumann \bcs}

If the fluctuating film boundary is coupled to the OP via Neumann \bcs, one obtains at long times the same expression for the variance $\bra \Delta R^2\ket_\infty$ as in \cref{eq_var_lateT}, but with $\Ncal$ based on \cref{eq_Q_Markov_Nbc} (with $R=0$).
Specifically, for $d=3$ and $\tauLG=0$ we obtain, by using the same estimates as in \cref{eq_var_lateT_d3_estim}, the long-time variance
\begin{subequations}
  \label{eq_var_lateT_d3_estim_Nbc}
    \begin{empheq}[left={\bra [\Delta R]^2\ket_\infty  =\empheqlbrace\,}]{align}  
	& \frac{1}{\kappa A} \left\{1 +  DA\left[ 0.85 + 0.042\times \ln(A^{1/2}/\varepsilon) + 0.020\times \frac{A}{L^2} \right]\right\}, &\text{(passive)} \label{eq_var_lateT_d3_estim_Nbc_pass} \\ 
	& \frac{1}{\kappa A}. &\text{(thermalized)} \label{eq_var_lateT_th_Nbc}
	\end{empheq}
\end{subequations}
\hspace{-0.1cm}Accordingly, also for Neumann \bcs the fluctuations of the CCF yield a subdominant contribution to $\bra [\Delta R]^2\ket_\infty$ for a passive boundary [\cref{eq_var_lateT_d3_estim_Nbc_pass}].

\subsection{Discussion}
\label{sec_msd_discussion}

\begin{figure}[t]\centering
    \subfigure[]{\includegraphics[width=0.31\linewidth]{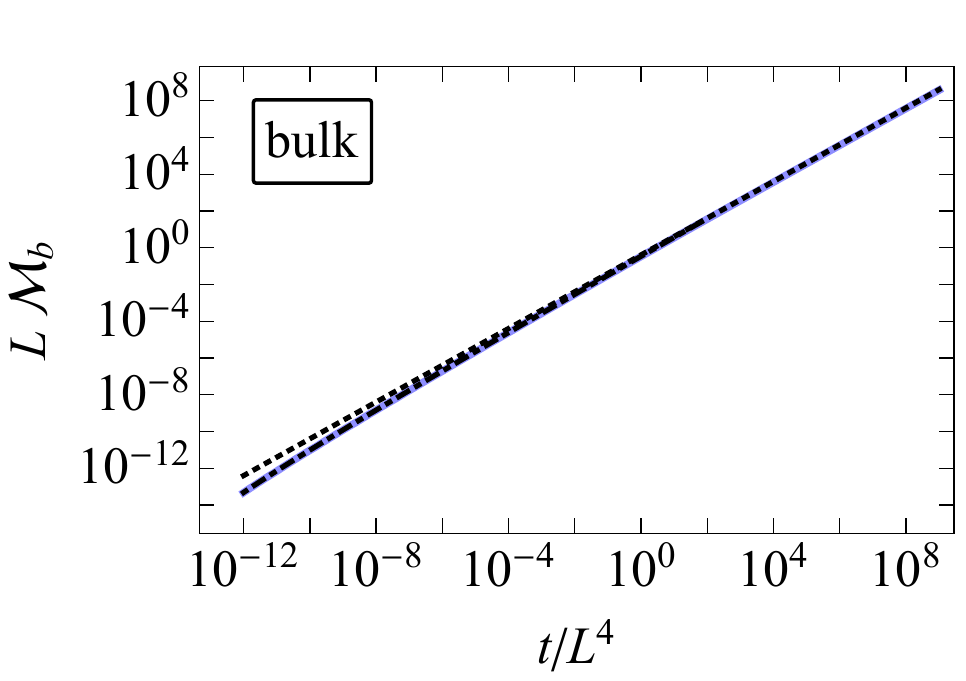} \label{fig_msd_bulk}}\quad 
    \subfigure[]{\includegraphics[width=0.31\linewidth]{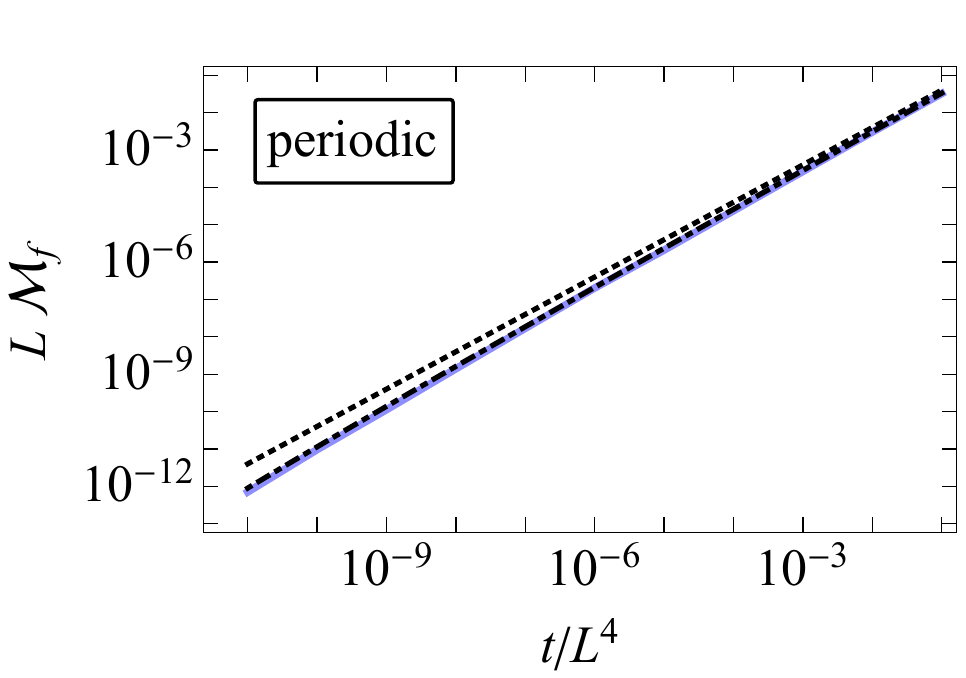}} \quad
    \subfigure[]{\includegraphics[width=0.31\linewidth]{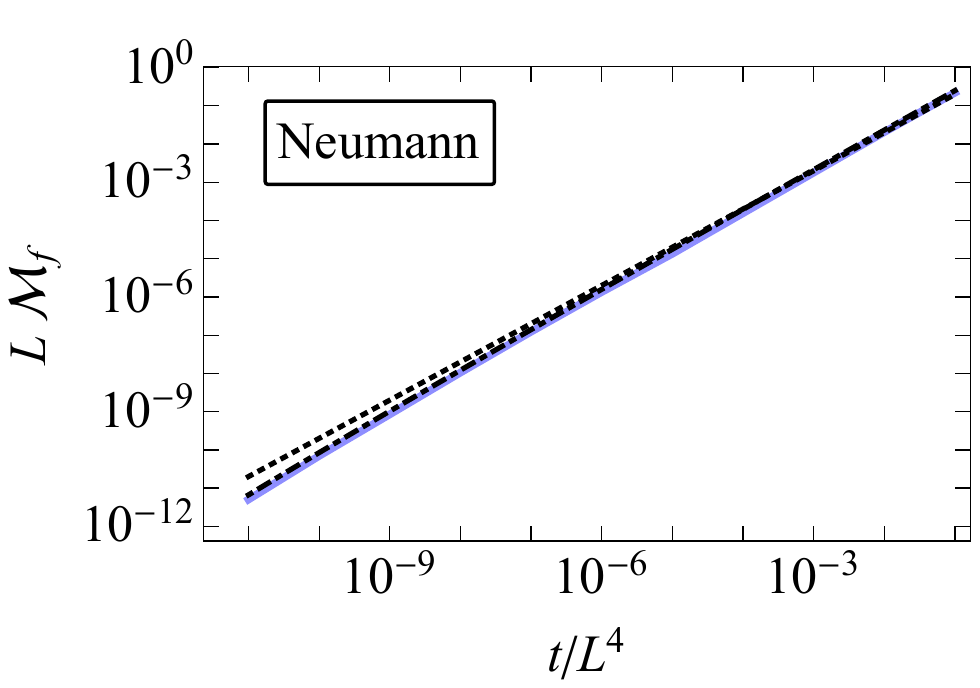}}
    \caption{Contributions $\Mcal_ {b,f}(t)$ [\cref{eq_msd_Pb,eq_msd_Pf}] to the mean-squared displacement of the movable film boundary at position $R(t)$ [see \cref{eq_msd_R_eval,eq_msd_R_blk}] in $d=3$ for (a) a bulk-like system ($R(t)\gg L$), (b) a film with periodic \bcs, and (c) a film with Neumann \bcs. The solid curves are the results of the numerical calculation of \cref{eq_msd_finalexpr} (with analogous expressions for the bulk and the film with Neumann \bcs, see the main text), while the dash-dotted curves (which practically overlap with the solid curves) represent the short-time scaling prediction reported in \cref{eq_msd_shortT_bulk,eq_msd_shortT_pbc,eq_msd_shortT_Nbc}. The dotted lines represent the function $0.37 \times t$, obtained from a numerical evaluation of \cref{eq_msd_lateT_bulk}, which describes the long-time behavior of $\Mcal_b$. The slight deviations from the behavior $\propto t$ at short times are due to the logarithmic correction in \cref{eq_msd_shortT_blk_d3corr,eq_msd_shortT_int_Nbc}. For illustrative purposes, we have used the value $\Lambda\approx 1000/\Lcal$ for the large momentum cutoff and $D=0.1/\Lcal^2$ for the bare diffusion constant. The curves depend only very weakly on the choices for these values. }
    \label{fig_msd_film}
\end{figure}

\Cref{fig_msd_film} illustrates the behavior of the critical-fluctuation induced contributions $\Mcal_{b,f}$ to the MSD [\cref{eq_msd_R_blk,eq_msd_R_eval}] of a movable film boundary in the adiabatic limit [$\chi=D\Lcal^2\ll 1$, see \cref{eq_adiab_param}], at bulk criticality ($\tauLG=0$) and in $d=3$.
\Cref{fig_msd_film}(a) shows the contribution $\Mcal_b$ [\cref{eq_msd_Pb}] due to the bulk pressure fluctuations, while Figs.~\ref{fig_msd_film}(b) and (c) illustrate the contributions $\Mcal_f$ [\cref{eq_msd_Pf}] stemming from the pressure fluctuations of a film with periodic or Neumann \bcs, respectively. 
The solid curve in \cref{fig_msd_film}(b) represents the numerical determination of the full expression in \cref{eq_msd_finalexpr} for periodic \bcs. The same presentation is given for the bulk and for a film with Neumann \bcs. These full results are  captured accurately by the asymptotic short-time approximations obtained in \cref{eq_msd_shortT_bulk,eq_msd_shortT_pbc,eq_msd_shortT_Nbc} (dash-dotted curves, overlapping with the solid ones).
We note that, in the bulk as well as in general at short times, the MSD of a passive and a thermalized boundary behave identically in a first approximation [see the discussion around \cref{eq_FPE_Markov_therm}].
According to \cref{eq_msd_R_blk,eq_msd_R_eval}, in the short-time limit the MSD in $d=3$ amounts  to
\beq \bra [\Delta R(t \ll \Lcal^4)]^2\ket \simeq 2 D t \left\{ 1  + \chi A\Lcal^{-2} \left[ a + b\, \ln(t\Lambda^4)\right]\right\},
\label{eq_msd_R_eval_lim}\eeq
where the constants $a$ and $b$ follow from \cref{eq_msd_shortT_blk_d3corr,eq_msd_shortT_int_Nbc} as $a_b=a\pbc \simeq 0.048$, $a\Nbc \simeq 0.068$, $b_b=b\pbc \simeq 0.011$, $b\Nbc \simeq 0.016$ for the respective \bcs as well as for the bulk (indicated by the super- and subscripts).
\Cref{eq_msd_R_eval_lim} applies under the additional assumption $t\ll 1/(DA\kappa)$, which is equivalent to $D t\ll \bra [\Delta R]^2\ket_\infty$ [obtained by using \cref{eq_var_lateT} in the adiabatic limit], as expected for the short-time regime.
Taking the largest length scale in the system to be $\Lcal\sim A^{1/2}$ (in $d=3$), the correction $\propto \chi$ in \cref{eq_msd_R_eval_lim} stemming from the fluctuations of the CCF is weak compared to the Brownian diffusion $2Dt$ induced by the thermal white noise.

At short times, the MSD behaves identically (up to constant prefactors) in the bulk and in proximity to another wall. This is physically expected, because the influence of the wall requires a certain amount of time to reach the moving boundary. 
The data in \cref{fig_msd_bulk} also cover the long-time regime of the bulk contribution $\Mcal_b$, which, according to \cref{eq_msd_lateT_bulk}, grows linearly upon increasing $t$ and depends logarithmically on a microscopic length scale.
While \cref{eq_msd_R_eval_lim} applies to $d=3$, for $d>3$, the contribution $\propto\chi$ to the MSD scales as $\Mcal_{b,f} \propto  t^{(7-d)/4}$ without a logarithmic correction [see \cref{eq_msd_shortT_bulk,eq_msd_shortT_pbc,eq_msd_shortT_Nbc}] and is dominated by the Brownian contribution $2Dt$ in \cref{eq_msd_R_eval_lim}. For $d\leq 2$, the correction terms $\Mcal_{b,f}$ depend stronger than linearly on time and additionally acquire logarithmic ($d=2$) or algebraic dependences ($d=1$) on the system size, indicating a potentially anomalous diffusion regime. The corresponding analysis is left for future studies.

At long times $t\gg 1/(DA\kappa)$, a movable boundary close to a wall explores the effective potential. Within the harmonic approximation, the position variance attains the value given by \cref{eq_var_lateT,eq_var_lateT_d3_estim_Nbc}.
Within the Markovian approximation, in fact an analytical expression for the full time dependence of the variance of the position follows from \cref{eq_FPE_Markov}: 
\beq \bra [\Delta R(t)]^2\ket = \bra [\Delta R]^2\ket_{\infty}\left[1-\exp(-2DA\kappa\, t) \right],
\label{eq_FPE_Markov_var}\eeq 
with $\bra [\Delta R]^2\ket_{\infty}$ given by \cref{eq_var_lateT} [together with \cref{eq_Knoise_Markov_var}, evaluated for $R=0$].
At short times $t\ll 1/(DA\kappa)$, \cref{eq_FPE_Markov_var} reduces to $\bra \left[\Delta R(t \ll 1/[DA\kappa])\right]^2\ket \simeq 2D t (1 + DA^2 \Ncal)$, which describes a standard Brownian motion with an effective diffusivity $D+(DA)^2\Ncal$.
In contrast, the non-Markovian character of the noise $\Delta\Kcal$ in general induces an algebraic short-time behavior [with possible logarithmic corrections, see \cref{eq_msd_shortT_bulk,eq_msd_shortT_pbc,eq_msd_shortT_Nbc,eq_msd_R_eval_lim}].
The MSD depends only weakly on these differences, because in both cases it is dominated, within the adiabatic approximation, by the Brownian diffusivity $D$.

\section{Summary, Conclusions, and Outlook}
\label{sec_summ}

In this study, we have analyzed the fluctuations of the critical Casimir force (CCF) acting on the boundary of a film with periodic, Neumann, or Dirichlet \bcs (\cref{fig_sketch}).
Our predictions are obtained within a time-dependent Ginzburg-Landau description of an equilibrium fluid system with a conserved order parameter (model B \cite{hohenberg_theory_1977}) and based on a rigorous definition of the instantaneous CCF \cite{dean_out--equilibrium_2010, kruger_stresses_2018, gross_surface-induced_2018, gross_dynamics_2019}.
We have considered a film consisting of a spatially fixed boundary and a parallel boundary which is either fixed (see \cref{sec_stat_press,sec_dyn_press}) or movable (see \cref{sec_fluct_boundary}). 
The movable boundary is coupled ``passively'' to the OP (see \cref{sec_dyn_model}) and is subject to the action of thermal white noise and the fluctuating CCF.
Due to the nature of the passive coupling, the boundary acts on the OP only by imposing \bcs, but not by exerting additional friction. By utilizing the fluctuation-dissipation relation, we have shown how this approximation can be improved in order to describe a boundary in thermal equilibrium (see \cref{sec_Markov_approx}).
We have distinguished between the motion in a bulk-like environment and the motion close to that boundary of the film which is kept fixed. 
In this latter case, an effective confinement potential arises due to the attractive mean part of the CCF and a short-ranged electrostatic repulsion.
This situation is analogous to a CCF-induced trapping of a colloid close to a solid wall, i.e., the inner boundary \cite{hertlein_direct_2008, paladugu_nonadditivity_2016, magazzu_controlling_2019}. 

Our results can be summarized as follows:
\begin{enumerate}
  \item The equilibrium variance of the CCF acting on a fixed boundary [\cref{eq_CCF_eqvar}] strongly depends on a microscopic cutoff length $\varepsilon$ and diverges in the continuum limit $\varepsilon\to 0$ [see \cref{eq_rms_CCF}], consistent with previous studies \cite{bartolo_fluctuations_2002, dantchev_critical_2004}. The cutoff $\varepsilon$ is determined by the length scale below which the continuum description of the fluid breaks down. The cutoff-dependent terms stem from both the film and the bulk pressure contribution.
 
  \item The dynamic correlation function of the CCF [\cref{eq_CCF_correl_dyn}] is finite at nonzero time differences and decays algebraically upon increasing time [see \cref{fig_PFilmCorrel}]. The short- and long-time scaling behaviors have been calculated analytically for general dimensionality $d$ and general values of the reduced temperature [see \cref{sec_dyncor_per,sec_dyncor_Neu}]. In a film in spatial dimension $d=3$, a weak logarithmic divergence is present at short times [see \cref{eq_Pf_earlyT_pbc_d3,eq_Pf_earlyT_Nbc_d3}], which is cut off by the system size.

  \item The motion of the film boundary is described by a Langevin equation [\cref{eq_langevin_bndry_full}] which takes into account the random forces emerging from both the fluctuating Casimir force and the thermal fluctuations due to the presence of the heat bath provided by the solvent. The Langevin equation can be solved perturbatively in the adiabatic limit, in which the order parameter field relaxes much faster than the fluctuating boundary. This solution is characterized by the following features:
  
  \begin{enumerate}
   \item In spatial dimension $d=3$ and at short times, the mean-squared displacement (MSD) $\bra\Delta R^2(t)\ket$ of the movable boundary increases linearly in time with a weak logarithmic correction [see \cref{eq_msd_R_eval_lim} and \cref{fig_msd_film}]. The contribution to the MSD stemming from the fluctuations of the CCF is typically small compared to the Brownian diffusion induced by the character of the fluid as a thermal bath. 
   
  \item In the case of a movable boundary which diffuses far from a wall, the Langevin equation yields a Brownian diffusive growth at long times, with a MSD $\propto t$ [see \cref{fig_msd_bulk}]. 
  
  \item In order to assess the long-time behavior of a boundary fluctuating close to a wall, we have derived a Fokker-Planck equation for the probability distribution of the boundary position, based on a Markovian noise approximation [see \cref{eq_FPE_Markov}]. At long times, this equation predicts a steady Gaussian distribution of the boundary position with the variance given by \cref{eq_var_lateT}. Within the passive boundary approximation, the variance acquires contributions from the fluctuations of the CCF [see \cref{eq_var_lateT_pass}]. By contrast, these contributions are absent in the case of a boundary in thermal equilibrium [see \cref{eq_var_lateT_th}]; the variance of its position is solely determined by the critical Casimir potential \cite{hertlein_direct_2008, gambassi_critical_2009,magazzu_controlling_2019}.  

  \end{enumerate}

\end{enumerate}

The formally divergent equilibrium variance of the CCF [see \cref{eq_rms_CCF}] can be considered as an artifact of the corresponding continuum field theory, which cannot be observed directly.
In fact, in previous studies of (critical) Casimir forces \cite{barton_fluctuations_1991, bartolo_fluctuations_2002}, it has been shown that this divergence problem is mitigated or even absent if time-averaged quantities are considered.
In the present study, we have instead directly linked the statistical properties of the CCF to experimentally observable quantities, such as the position of a film boundary.
Our analysis renders a position variance which is finite and reduces to the variance of a Brownian particle in an effective potential set by the mean CCF.
This finding is consistent with previous experiments on CCFs in wetting films \cite{garcia_critical_1999, ganshin_critical_2006, fukuto_critical_2005, rafai_repulsive_2007} or in colloidal systems \cite{hertlein_direct_2008, gambassi_critical_2009}. Heuristically, this situation is analogous to the fact that observable quantities obtained from a standard Langevin equation are usually finite, despite the random force being correlated $\propto \delta(t-t')$, which formally diverges in the static limit $t\to t'$.
In the present case, the force correlations diverge algebraically (see \cref{fig_PFilmCorrel}) but are still integrable in the sense of the MSD (see the discussion in \cref{sec_integr}).
Moreover, due to the limited temporal resolution of a measurement device, experimental measurements are typically not sensitive to fluctuations of the CCF at frequency scales pertaining to the relaxation of the short-wavelength modes \cite{bartolo_fluctuations_2002}. 

Our analysis of the dynamics of the boundary is based on two central assumptions: first, the moving boundary acts on the OP only via the imposed \bcs, not via coupling terms in the equation of motion (\emph{passive boundary approximation}) \cite{demery_perturbative_2011,dean_diffusion_2011,gross_dynamics_2021}. While, within this approximation, detailed balance is broken and the system is inherently out of equilibrium, we have also shown how to re-establish detailed balance at the level of the Fokker-Planck equation for the position of the boundary [see \cref{eq_FPE_Markov_therm}]. Second, the relaxation dynamics of the boundary is much slower than the one of the order parameter; this is fulfilled in typical experiments \cite{magazzu_controlling_2019} and also ensured in our study because we have assumed a finite, albeit macroscopically large, system size.
Within the passive boundary approximation, the actual Casimir potential---which is related only to the mean CCF---cannot be simply inferred by assuming a standard Boltzmann distribution of the position \cite{gambassi_critical_2009}, because OP fluctuations render additional contributions to the variance which can vary in space [see \cref{eq_Knoise_Markov_var}]. Asymptotically, in the adiabatic limit these contributions are negligible.

Concerning future research, a systematic derivation of the dynamical equation from a more fundamental model, beyond the passive boundary approximation, would be desirable.
It would also be interesting to extend the present analysis towards the dynamics of a spherical colloidal particle which has a finite radius, is immersed in a critical solvent, and floats near a wall \cite{magazzu_controlling_2019,vasilyev_nonadditive_2018}. 
It is justified to expect a wealth of additional effects due to hydrodynamic interactions, the non-planar geometry, and the strong preferential adsorption of the OP at the surfaces \cite{furukawa_nonequilibrium_2013,fujitani_fluctuation_2016,yabunaka_drag_2020}.

\begin{acknowledgments}
AG acknowledges financial support from MIUR PRIN project ``Coarse-grained description for non-equilibrium systems and transport phenomena (CO-NEST)'' n. 201798CZL.
\end{acknowledgments}

\appendix

\section{Equilibrium distribution of a movable boundary}
\label{app_equil_PDF}

Here, we discuss the probability distribution of the position of a movable boundary in thermal equilibrium with a critical fluid in a half-space \cite{dean_out--equilibrium_2010,gross_surface-induced_2018}. 
The Hamiltonian of this system is given by $\Fcal(R;[\phi])= \int_V \d^d r\, \Hcal(\rv,R, \phi(\rv), \nabla\phi(\rv), \tauLG)$ [see \cref{eq_freeEn}], where the position $R$ of the boundary is regarded as an additional degree of freedom, which emerges via the surface contribution $\Hcal_s$ [\cref{eq_Hden_s}].
Accordingly, the joint probability distribution is given by [see \cref{eq_eq_dist}] 
\beq P\st{eq}([\phi],R) \sim e^{-\Fcal(R; [\phi])/T}.
\eeq 
The equilibrium probability distribution of $R$ follows by integrating over $\phi$:
\beq 
P\st{eq}(R) = \frac{\Zcal(R)}{\Zcal_0} ,\qquad \Zcal(R) = \int \Dcal\phi\, e^{-\Fcal(R;[\phi])/T}, \qquad \Zcal_0 = \int_0^\infty \d R\, \Zcal(R).
\label{eq_Ps_R}\eeq 
The mean equilibrium CCF $\bra\Kcal\ket$ (per area $A$ and temperature $T$) acting on the boundary [which in \cref{eq_CCF_Tzz} is expressed in terms of the averaged stress tensor] can be equivalently obtained from $\Zcal(R)$ \cite{dean_out--equilibrium_2010}:
\beq A \bra\Kcal(R)\ket_\phi =  \frac{\d}{\d R} \ln \Zcal(R).
\label{eq_CCF_freeEn}\eeq 
Upon introducing the critical Casimir potential $V_C(R)$ via $\bra\Kcal(R)\ket_\phi = -\d V\st{C}(R) / \d R$, and by using \cref{eq_CCF_freeEn}, the distribution $P\st{eq}$ in \cref{eq_Ps_R} can be expressed  as
\beq P\st{eq}(R) = \frac{1}{\Zcal} e^{-A V_C(R)}.
\label{eq_Ps_R_Vc}\eeq 
Accordingly, if an inclusion is represented in terms of a Hamiltonian interaction with the fluid (as, e.g., in Refs.\ \cite{demery_perturbative_2011,dean_diffusion_2011,gross_dynamics_2021}), the fluctuations of its position $R$ are actually controlled by the potential of mean Casimir force.

%==================================================================================================

%\bibliography{bibliography}

\end{document}